\documentclass[10pt,a4paper]{article}
\usepackage[margin=1.5in]{geometry}

\usepackage[english]{babel}
\usepackage[utf8x]{inputenc}
\usepackage[T1]{fontenc}
\usepackage{amsmath}
\usepackage{amsfonts}
\usepackage{amssymb}
\usepackage{amsthm}

\usepackage{lipsum}
\usepackage{amsfonts}
\usepackage{graphicx}
\usepackage{epstopdf}

\usepackage{algorithm}
\usepackage[noend]{algpseudocode}

\newcommand{\email}[1]{\protect\href{mailto:#1}{#1}}

\usepackage{gensymb}
\usepackage{graphicx}
\usepackage{hyperref}
\usepackage{listings}
\usepackage{scrextend}
\usepackage{stix}
\usepackage{tikz}

\usepackage{xfrac}
\usepackage{verbatim}
\usepackage{cprotect} 
\usepackage{bbm} 
\usepackage{amsbsy}
\usepackage{caption} 
\usepackage{xcolor}
\usepackage{scalerel}
\usepackage{url}
\usepackage{float}

\theoremstyle{definition}
\newtheorem{definition}{Definition}
\newtheorem{proposition}{Proposition}
\newtheorem{corollary}{Corollary}

\theoremstyle{plain}
\newtheorem{lemma}{Lemma}

\theoremstyle{remark}
\newtheorem{remark}{Remark}

\usepackage{amsopn}

\usepackage{draftwatermark}
\SetWatermarkText{Accepted version}
\SetWatermarkScale{.5}
\SetWatermarkColor[rgb]{.9,.9,.9}

\def\C{\ensuremath\mathbb{C}}
\def\R{\ensuremath\mathbb{R}}
\def\S{\ensuremath\mathcal{S}}
\def\Sig{\ensuremath\Sigma}
\def\N{\ensuremath\mathbb{N}}

\def\L{\ensuremath\mathbb{L}}
\def\H{\ensuremath\mathcal{H}}

\def\D{\ensuremath\mathbf{D}}
\def\A{\ensuremath\mathbf{A}}

\def\alph{\ensuremath \boldsymbol{\alpha}}
\def\Phh{{\boldsymbol{\Phi}}}

\def\Zbf{\ensuremath\mathbf{Z}}
\def\Ubf{\ensuremath\mathbf{U}}
\def\Vbf{\ensuremath\mathbf{V}}
\def\Dbf{\ensuremath\mathbf{D}}

\def\arr{\ensuremath\overrightarrow}
\def\r{\ensuremath\arr{r}}

\def\z{\ensuremath\mathbf{z}}
\def\w{\ensuremath\mathbf{w}}
\def\u{\ensuremath\mathbf{u}}
\def\v{\ensuremath\mathbf{v}}
\def\x{\ensuremath\mathbf{x}}
\def\y{\ensuremath\mathbf{y}}
\def\cbf{\ensuremath\mathbf{c}}
\def\d{\ensuremath\mathbf{d}}
\def\de{\ensuremath\boldsymbol{\delta}}
\def\rbf{\ensuremath\mathbf{r}}
\def\qbf{\ensuremath\mathbf{q}}

\DeclareMathOperator*{\argmax}{arg\,max}
\DeclareMathOperator*{\argmin}{arg\,min}
\newcommand\herH[2]{\ensuremath ( #1 ~|~ #2 )_{\H}}

\newcommand\herPhi[2]{\ensuremath ( #1 ~|~ #2 )_{\Phh}}

\newcommand\shape[1]{\ensuremath \left[\Pi_{\S}(#1)\right]} 

\author{Anna Song \thanks{Department of Mathematics and Applications, \'Ecole Normale Sup\'erieure (ENS), Paris, France 
		(\email{anna.song.maths@gmail.com}).}
	\and Virginie Uhlmann \footnotemark[5] \thanks{European Bioinformatics Institute (EMBL-EBI), Cambridge, UK (\email{uhlmann@ebi.ac.uk})}
	\and Julien Fageot \footnotemark[5] \thanks{Signals, Information, and Networks Group (SING), Harvard University, Cambridge, USA \newline (\email{julienfageot@fas.harvard.edu}).}
	\and Michael Unser \thanks{Biomedical Imaging Group (BIG), \'Ecole Polytechnique F\'ed\'erale de Lausanne (EPFL), Lausanne, Switzerland}
}

\title{Dictionary Learning for \\ Two-Dimensional Kendall Shapes
	\thanks{
		This work was supported by \'Ecole Normale Sup\'erieure and \'Ecole Polytechnique F\'ed\'erale de Lausanne. It was also funded by Swiss National Science Foundation, Grant \url{200020_162343 / 1}. It was partly supported by EMBL core fundings and the Swiss National Science Foundation with grant agreement \url{P2ELP2_181759}.}    }

\begin{document}

\maketitle

\begin{abstract}
We propose a novel sparse dictionary learning method for planar shapes in the sense of Kendall, namely configurations of landmarks in the plane considered up to similitudes. Our shape dictionary method provides a good trade-off between algorithmic simplicity and faithfulness with respect to the nonlinear geometric structure of Kendall's shape space. Remarkably, it boils down to a classical dictionary learning formulation modified using complex weights. Existing dictionary learning methods extended to nonlinear spaces either map the manifold to a reproducing kernel Hilbert space or to a tangent space. The first approach is unnecessarily heavy in the case of Kendall's shape space and causes the geometrical understanding of shapes to be lost, while the second one induces distortions and theoretical complexity. Our approach does not suffer from these drawbacks. Instead of embedding the shape space into a linear space, we rely on the hyperplane of centered configurations, including pre-shapes from which shapes are defined as rotation orbits. In this linear space, the dictionary atoms are scaled and rotated using complex weights before summation. Furthermore, our formulation is more general than Kendall's original one: it applies to discretely-defined configurations of landmarks as well as continuously-defined interpolating curves. We implemented our algorithm by adapting the method of optimal directions combined to a Cholesky-optimized order recursive matching pursuit. An interesting feature of our shape dictionary is that it produces visually realistic atoms, while guaranteeing reconstruction accuracy. Its efficiency can mostly be attributed to a clear formulation of the framework with complex numbers. We illustrate the strong potential of our approach for the characterization of datasets of shapes up to similitudes and the analysis of patterns in deforming 2D shapes.
\end{abstract}

\begin{center}
\end{center}

\begin{center}
	\textbf{AMS}
	94A12, 65J22, 51M99, 65D18, 65D05
\end{center}

\section{Introduction}
\label{sec:intro}
Shape analysis is highly relevant to biomedical imaging and computer vision. Among many other applications, it may be deployed to retrieve the main features from a collection of shapes, compare shapes as in morphometrics \cite{bennett2012,drake2015}, classify or recognize objects \cite{costa2010,ovsjanikov2009}, describe the dynamics of a moving object or organism \cite{stephens2008,buckingham2008}, or study the distribution of data in a shape space \cite{dryden2016,small2012} relatively to a mean shape \cite{stegmann2002}.
Depending on the application,
the concept of shape has different meanings \cite{costa2010,zhang2004,loncaric1998}. Traditionally, shapes are handled as vectors $\x_1,...,\x_K \in \R^d$ and, for instance, correspond to silhouettes of objects outlined with landmarks. Standard signal analysis tools such as principal component analysis (PCA) can then be used to find the main modes of variation in the dataset of shapes \cite{bookstein1997,dryden2016,stegmann2002,stephens2008}.

Alternatively, one can improve results obtained with PCA by using more refined tools, such as sparse dictionary learning  \cite{elad2010,kreutzdelgado2003,mairal2012,mairal2009}. Given a \textit{dictionary} of representative elements referred to as \textit{atoms}, a signal is reconstructed from a \textit{sparse} linear combination of them that minimizes the approximation error. This \textit{sparse coding} is motivated by the assumption that natural signals are sparse \cite{mairal2014,olshausen1996,donoho2006,unser2014,fageot2015}. The notion of sparsity has already proven its importance in
an extensive range of problems, from image denoising and signal recovery to recognition and classification \cite{mairal2014,elad2010,candes2006,wright2009}. Moreover, when the dictionary is learned from the data, significant improvements can be made on the signal reconstruction \cite{olshausen1996,elad2006,mairal2012}, leading to the so-called sparse dictionary learning approach, classically formulated as
\begin{equation}\label{eq:dico_classical}
\inf\limits_{\D,\A}  \sum_{k=1}^K \|\x_k - \D \alph_k\|^2 + \lambda \text{Sp}(\alph_k),
\end{equation}
where $\D = (\d_1,...,\d_J) \in \R^{d \times J}$ is the dictionary and $\A = (\alph_1,...,\alph_K) \in \R^{J \times K}$ contains the \textit{weights} used to reconstruct $\x_k$. A bounding constraint on the atoms, $\|\d_j\| \leq 1$, is added to ensure the well-posedness of the minimization problem. Sparsity is promoted by the term $\lambda \text{Sp}(\alph_k)$ which penalizes nonzero coefficients in the weights, with $\lambda > 0$ a parameter. Often, one relies on $\text{Sp}(\alph_k) = |\alph_k|_0$ or $\text{Sp}(\alph_k) = |\alph_k|_{1}$, which denote the $\ell_0$ constraint and the $\ell_1$ norm, respectively. When $\hat{\D}$ and $\hat{\A}$ approximately optimize \eqref{eq:dico_classical}, $\hat{\D} \hat{\alph}_k$ is the \textit{reconstruction} of the original data $\x_k$. \newline

In some applications, the data are pre-processed in order to discard the influence of uninformative features such as the specific position, size, and orientation of the silhouettes. As in Procrustes analysis \cite{dryden2016,stegmann2002}, original silhouettes are scaled, translated, and rotated so as to optimally match a reference silhouette, typically the mean of the dataset.
However, for data with high variability, this approach sometimes fails to produce visually \textit{interpretable} representative elements (modes or atoms) because, after alignment, silhouettes are still not position-, scale-, and orientation-independent.
To address this, we should handle data as \textit{shapes} in the sense of Kendall \cite{kendall1977,kendall1984}. By definition, two geometric objects have the same shape if they are equivalent up to (direct) \textit{similitudes} (\textit{i.e.}, up to translation, scaling, and rotation).
Shapes are hence considered as nonlinear objects. Our work is motivated by the lack of a simple and efficient sparse dictionary learning method dedicated to Kendall's shape space. Such an analysis would be truly \textit{invariant to similitudes}, and benefit from both the efficiency of sparse dictionary learning and a valuable shape analysis framework \cite{kendall1977,kendall1984,dryden2016,srikla2016,stegmann2002,jayasumana2013,bentanfous2018}.

An additional interesting feature of our approach is that it is suitable for the analysis of shapes defined from discretely-defined configurations of landmarks as well as continuously-defined curves.
In Kendall's original formulation, the shape space is built upon \textit{configurations of landmarks} in $(\R^d)^N$ considered up to similitudes \cite{kendall1984,dryden2016,srikla2016}.
Here, we extend Kendall's shape space to \textit{interpolating curves} linearly generated by finitely many basis functions.
This extended framework applies in particular to \textit{spline curves}
generated by piecewise polynomials.
They constitute a common representation for parametric curves \cite{unser1993,brigger2000,gonzalo2013,uhlmann2016,uhlmann2017,schmitter2018}. Our work is hence related to \cite{schmitter2018}, which relies on the isometry between spline curves and configurations of landmarks to apply dictionary learning after pre-alignment. \newline

\paragraph{Related works}

Standard dictionary learning methods cannot be straightforwardly extended to Kendall's shape space because the latter is not a vector space but a Riemannian manifold. The difficulty resides in the \textit{nonlinear} geometric structure of this space, in which a meaning must be given to linear combinations of atoms. Until recently, most dictionary learning approaches were devoted to data lying in linear spaces. In the past years, however, a few works have focussed on the extension of sparse dictionary learning to nonlinear spaces such as the Grassmann manifold, the manifold of symmetric positive definite (SPD) matrices, Kendall's shape space (as in the two works \cite{bentanfous2018,jayasumana2013} that are closely related to ours), and more general Riemannian manifolds. This can be achieved in two ways: either by mapping the manifold to a Hilbert space, typically a reproducing kernel Hilbert space (RKHS) \cite{harandi2012,caseiro2012,li2013,jayasumana2013,harandi2015,harandi2016}; or by projecting the data onto a tangent space, once at a reference point, or iteratively at multiple points \cite{cetingul2009,yuan2009,guo2010,xie2013,vemulapalli2014,anirudh2015,huang2017,bentanfous2018}. Both approaches flatten the nonlinear space by mapping it to a linear one, so as to make weighted sums of atoms possible.

On the one hand, the mapping of Kendall's space to a RKHS through the kernel trick, as done in \cite{jayasumana2013}, follows a classical procedure and enables, after mapping, the use of several established techniques. However, it looses the simple structure of the underlying shape space because the mapping is not explicit. On the other hand, projecting the dataset on one tangent space unfaithfully represents the original distances outside of a neighborhood around the pole \cite{caseiro2012,harandi2012}. In \cite{bentanfous2018}, an analysis with multiple projections instead is proposed by the authors. They adapt to Kendall's 3D shape space a sparse dictionary learning method first proposed in \cite{xie2013} for general Riemannian manifolds, with the analysis of trajectories of shapes for 3D action recognition as targeted application.

Although the approach of \cite{xie2013,bentanfous2018} has proven its success for action recognition tasks, some of its characteristics are less relevant here. It requires one to fold and unfold the manifold several times on tangent spaces using $\log_{\x_k}$ and $\exp_{\x_k}$ mappings. Most importantly, an affine constraint is added in order to ensure a nontrivial solution, but this requirement modifies the original problem, as pointed out by \cite{harandi2013}. In addition, the original shapes $\x_k$ themselves are still needed in their reconstruction, which is undesirable here since the goal is to reconstruct a dataset from the dictionary and weights only. As we do not intend to tackle the same applications, our method should be considered as an alternative for learning shapes. \newline

\paragraph{Overview of our method}
In contrast to all previous approaches, the sparse dictionary learning method that we propose is mathematically simple while remaining faithful to the nonlinear structure of the shape space. We avoid kernel methods because they provide no additional geometrical understanding to the problem and are too sophisticated in regard of the simple structure of the shape space. We do not rely on tangent projections either, thereby avoiding both distortions and theoretical complexity.

Instead, the key idea of our approach can be summarized as follows. Kendall's shape space is the quotient of a \textit{pre-shape sphere} $\S$ by the group of rotations. The manifold $\S$ is the sphere inside the linear space composed of all centered configurations. We compute all linear combinations in this vector space. Our atoms are \textit{pre-shapes} in $\S$ (\textit{i.e.}, centered and normalized configurations). We use complex numbers in the linear combination so as to scale and rotate them. We then add them together to obtain a configuration whose shape is close to that of the original one.
This proximity is measured through a particular metric in Kendall's shape space. For the classical full Procrustes distance, our method simply leads to the optimization problem
\begin{equation} \label{eq:dico_simple0}
\inf \limits_{ \D, \A : \substack{\hskip-.5cm \d_j \in \S \\ |\alph_k|_0 \leq N_0 } } \sum_{k=1}^K \left|\z_k - \D \alph_k \right|_\Phh^2,
\end{equation}
where $|\cdot|_\Phh$ is an $\ell^2$ norm specific to the representation (landmarks or interpolating curves), $\alph_k$ are complex weights, and the data $\z_k$ and atoms $\d_j$ are (complex) pre-shapes. Our work therefore proposes a natural extension of the standard dictionary learning method with a strong theoretical justification.\newline

\paragraph{Contributions}

\begin{enumerate}
	\item We extend Kendall's original framework to continuously-defined interpolating curves. For this purpose, we put forward the concept of \textit{configuration}, which indifferently represents a discrete object (configuration of landmarks) or a continuous one (interpolating curve).
	We thus extend and embed the work from \cite{schmitter2018} inside the more general (and appropriate) framework of Kendall's shape analysis.
	\item
	Our entire approach of the 2D shape space is formulated in terms of complex numbers and Hermitian inner products. The geometrical interpretation of complex numbers provides a clear understanding of the framework.
	\item Our main contribution consists in a simple and efficient sparse dictionary learning method, that we call \textit{2D Kendall Shape Dictionary} (2DKSD), dedicated to the analysis of 2D shapes in the sense of Kendall. Our approach provides faithful reconstructions with respect to the nonlinear geometric structure of Kendall's shape space, while remaining mathematically light. Using the full Procrustes distance \cite{dryden2016} to compare the original and reconstructed shapes, 2DKSD boils down to \textit{a simple and nearly classical dictionary learning formulation} \eqref{eq:dico_simple0} that relies on complex weights instead of real ones. It allows one to scale and rotate the atoms inside the weighted sum individually for each data point, instead of aligning the dataset to a reference shape as a pre-processing step.
	\item Our implementation of this simple formulation is an adaptation of the algorithm used in the \verb|SPAMS| software \cite{mairal2014} to the Hermitian framework, and is freely available online.\footnote{https://github.com/ansonang3/2DKSD} More precisely, we combine the method of optimal directions \cite{engan1999} to an order recursive matching pursuit \cite{cotter1999} with a Cholesky-based optimization.
	\item Thanks to the complex setting, our method provides better reconstruction accuracy than approaches relying on the real setting. The atoms of 2DKSD are also visually more realistic and similar to shapes present in the original dataset.\newline
\end{enumerate}

\paragraph{Outline of the article}

In Section \ref{sec:configurations}, we introduce the notion of planar configurations $\z$, and describe the action of similitudes over them.
Section \ref{sec:kendall} is devoted to Kendall's space of 2D shapes, reformulated for general configurations. We briefly recall the structure of the shape space and its three classical metrics. Then, in Section \ref{sec:shapedico}, we expose our main contribution: a sparse dictionary learning method dedicated to the analysis of 2D Kendall shapes, which for a well-chosen shape metric leads to a nearly standard formulation with complex weights \eqref{eq:dico_simple0}. We expose the implementation of 2DKSD in Section \ref{sec:algorithms}. Finally, we validate our approach in Section \ref{sec:results} by experimenting on shapes extracted from real image datasets, before concluding in Section \ref{sec:ccl}.

\begin{table}[h]
	\caption{\textbf{Spaces and elements.}}
	\label{tab:notations}
	\centering
	\begin{tabular}{|c|c|c|c|}
		\hline
		\textbf{element} & \textbf{notation} & \textbf{space} & \textbf{real dim}\\
		general configuration & $\z$ & $\C^N$ & $2N$ \\
		shift configuration & $\u$ & $\C^N$ & $2N$ \\
		centered configuration & $\z_0 = \z - \frac{\u^* \Phh \z}{|\u|^2_\Phh} \u$ & $(\C \u)^{\perp_\Phh} \simeq \C^{N-1}$ & $2N - 2$ \\
		pre-shape (centered, normalized) & $\Pi_{\S}(\z) = \frac{\z_0}{|\z_0|}_\Phh$ & $\S = \{|\z|_\Phh = 1\} \cap (\C \u)^{\perp_\Phh} $ & $2N-3$ \\
		shape (pre-shape up to rotations) & $[\z]$ & $\Sig = \S / U(1)$ & $2N-4$ \\
		\hline
	\end{tabular}
\end{table}

\begin{figure}[h]
	\centering
	\includegraphics[width=.7\linewidth]{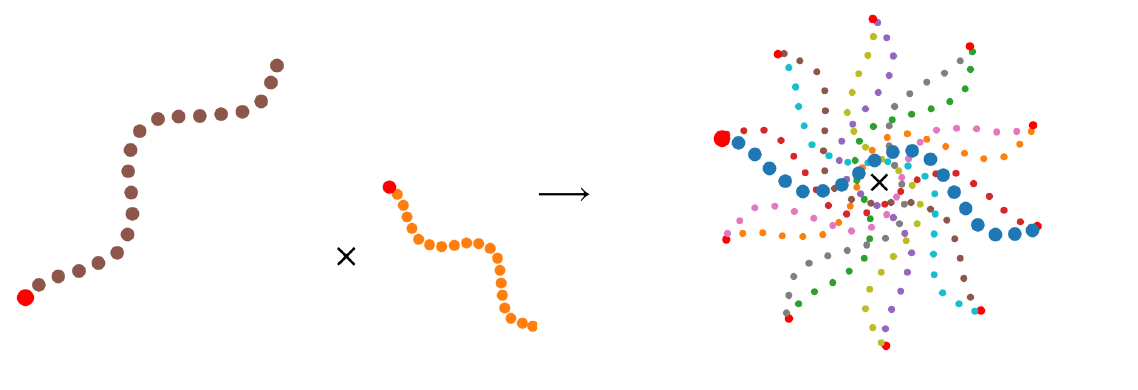}
	\caption{\small \textbf{The notion of shape in the sense of Kendall: non-degenerate configurations up to similitudes.} $\z_1 \sim \z_2$ (left) are non-degenerate configurations, equivalent up to direct similitudes. The cross indicates the origin. Hence, they share the same shape $[\z]$ (right), which can also be identified to a corresponding pre-shape (blue) up to rotations.}
	\label{fig:notion_shape}
\end{figure}

\begin{figure}[H]
	\centering
	\includegraphics[width=\linewidth]{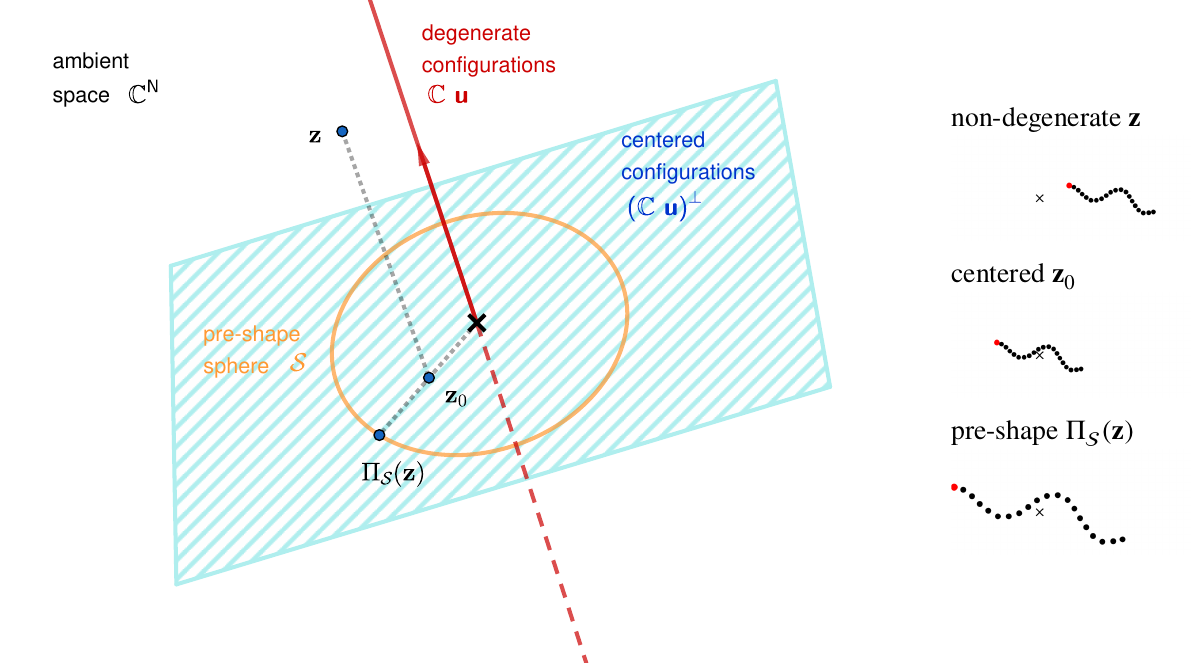}
	\caption{\small \textbf{Illustration of the spaces and of the ``pre-shaping'' operations.} Atoms are pre-shapes that are linearly combined inside the (complex) hyperplane of centered configurations.\textsuperscript{a}}
  \footnotesize \flushleft \textsuperscript{a}  For the purpose of illustration, elements are drawn in analogy to a real setting.
	\label{fig:preshaping}
\end{figure}

\section{Configurations in the plane and similitudes}
\label{sec:configurations}

\subsection{A fundamental example: landmarks (discrete)}

Traditionally, the silhouette of a 2D object is sampled by an \textit{ordered} set of points, called \textit{landmarks}, that often mark salient features of the boundary. The object itself is then represented by this collection of $N$ points, which we call a \textit{configuration of $N$ landmarks}, and is an element of $\C^N$ denoted by the bold letter $\z$. We are now interested in how (direct) similitudes transform configurations. In the sequel, the word \textit{similitudes} refer to direct similitudes.

It is elementary but important to observe that the action of a similitude in the plane, denoted by $(a,b) \in \C^* \times \C$ (where $\C^* := \C \setminus \{0\}$), has a nice expression using complex numbers
\begin{equation}
 \forall z \in \C, \qquad (a,b) ~\odot~ z = a z + b,
\end{equation}
where the multiplication by $a$ applies to $z$ a scaling of modulus $|a|$ and a rotation of angle $\arg a$, while the addition with $b$ encodes the effect of a translation. We recall that $\arg$ is the complex argument: for $z$ in $\C$, $z = |z| e^{\mathrm{i} \arg(z)}$, with $\arg(z) \in [0,2 \pi)$.
Similitudes form a group $(\mathcal{G},\circ)$ whose composition law is given by $(a',b') ~\circ~ (a,b) = (a'a, a'b + b').$

Similarly, the action of similitudes on configurations of landmarks also has a natural expression
$$\forall \z \in \C^N, \qquad (a,b) ~\odot~ \z = a \z + b \u,$$
where $\u = (1,...,1)$ is the configuration collapsing to $1$. In Figure \ref{fig:notion_shape} (left), the two configurations $\z_1$ and $\z_2$ are obtained from each other by a similitude. We say that they are \textit{equivalent up to similitudes}.

\subsection{Interpolating curves (continuous)}
\label{subsub:splines}
We now move to a more involved concept of configuration: interpolating curves, that are used as an alternative to landmarks for representing silhouettes. In the plane, \textit{interpolating curves} with $N$ degrees of freedom are built by interpolating $N$ complex coefficients $\z[0],...,\z[N-1]$ using linearly independent \textit{basis functions} $\phi_n \in \L^2([0,1],\R)$, $n = 0,...,(N-1)$. They have the general expression
\begin{equation}
\label{eq:def_spline_curve}
\forall t \in [0,1], \quad r(t) = \sum_{n = 0}^{N-1} \z[n] \phi_n(t).
\end{equation}
If $r(0) = r(1)$, then $r$ is a \emph{closed interpolating curve}, otherwise it is an \emph{open interpolating curve}.
The vector $\z \in \C^N$ is called ``the'' \textit{control vector} of $r$ (as we show below, $\z$ is uniquely defined).

The $\phi_n$ are often taken to be continuous, as in the case of \textit{interpolating spline curves}, or simply \textit{spline curves}, for which the $\phi_n$ are piecewise-polynomial.
Interpolating curves are considered as continuously-defined objects, in opposition to discretely-defined landmarks. Yet, they are in fact intermediate objects between the continuous and discrete settings. As illustration, we describe the construction of interpolating curves relying on cubic B-spline interpolation in Appendix \ref{sub:appendix_splines}.\newline

\paragraph{Isometry}
Interpolating curves are elements of the space of square-integrable curves\\$\H := \L^2([0,1],\C)$.
We endow $\H$ with the standard Hermitian inner product
\begin{equation}
\herH{r}{s} := \int_{0}^{1} \bar{r} ~ s = \int_{0}^{1} (r_1 s_1 + r_2 s_2) ~ + \mathrm{i} \int_{0}^{1} (r_1 s_2 - r_2 s_1), \qquad |r|_{H} = \sqrt{\herH{r}{r}},
\end{equation}
where $r = r_1 + \mathrm{i} r_2 \in \H$, and similarly for $s$.
Let $\Gamma$ denote the linear map
\begin{equation}
\displaystyle \Gamma : \begin{array}{ccc}
\C^N & \to & \H \\
\z & \mapsto & \sum_{n = 0}^{N-1} \z[n] \phi_n
\end{array}
\end{equation}
and $\Phh \in \R^{N \times N}$ the Gram matrix of the basis functions $\phi_n$, defined by
\begin{equation}
\Phh[n,m] := \langle \phi_n, \phi_m \rangle_{\L^2([0,1],\R)} = \int_0^1 \phi_n(t) \phi_m(t) \,\mathrm{d}t.
\end{equation}
The matrix $\Phh$ is real-valued, symmetric, positive-definite. 
Let us endow $(\C^N,\Phh)$ with the Hermitian inner product (or shortly, \textit{Hermitian product}) associated to $\Phh$ : the product of $\z,\w \in \C^N$ is $\herPhi{\z}{\w} = \z^*\Phh\w$, where $\z^* := \bar{\z}^T$ refers to the conjugate transpose of $\z$. The corresponding norm is then $|\z|_\Phh := \sqrt{\z^*\Phh\z}$. In Appendix \ref{sub:appendix_hermitian}, we detail how the Hermitian product is related to the real scalar product. As a recall of the conjugate symmetry, please note that $\z^*\Phh\w = \overline{\w^*\Phh\z}$. 

\begin{proposition}[Isometry] \label{prop:isometry}
	The linear map $\Gamma$ is an isometry from $(\C^N,\Phh)$ to $(\H,\herH{~}{~})$, in the sense that
	\begin{equation} \label{eq:herpro}
	\forall~ \z,\w \in \C^N, \quad \herH{\Gamma(\z)}{\Gamma(\w)} = \herPhi{\z}{\w}.
	\end{equation}
	Therefore, interpolating curves form an $N$-dimensional subspace $\Gamma(\C^N) \subset \H$ isometric to $(\C^N,\Phh)$.
\end{proposition}
\begin{proof}
	We develop the product $\herH{\sum_{n = 0}^{N-1} \z[n] \phi_n }{\sum_{m = 0}^{N-1} \w[m] \phi_m}$ to obtain \\
	$\sum_{n,m} \overline{\z[n]} \w[m] \langle \phi_n, \phi_m \rangle_{\L^2([0,1],\R)}$, which is exactly $\z^* \Phh \w$.
\end{proof}

As a consequence, an interpolating curve $r \in \Gamma(\C^N)$ is completely identified to its control vector $\z = \Gamma^{-1}(r) \in \C^N$. 
This result is important for the extension of Kendall's theory to interpolating curves\footnote{The question of defining a good notion of curve shape and, in particular, a suitable representation for the curve (such as the square-root velocity (SRV) representation \cite{srikla2011}), is outside the scope of this article. We refer interested readers to \cite{srikla2016}.} and ensures that they can be identified to configurations.\newline

\paragraph{Action of similitudes over interpolating curves}
The action of any similitude $(a,b) \in \C^* \times \C$ on curves in $\H$ has once again a nice formulation
\begin{equation}
\forall r \in \H, \qquad (a,b) ~\odot~ r = ar + b \mathbbm{1},
\end{equation}
where $\mathbbm{1}$ refers to the constant function $\forall t \in [0,1], ~t \mapsto 1$ in $\H$.
Suppose that $\mathbbm{1} \in \Gamma(\C^N)$, namely, that constant curves are generated by the same basis functions $\phi_n$. Under this assumption, it is legitimate to define $\u := \Gamma^{-1}(\mathbbm{1})$.
The action of similitudes is induced on $\C^N$ according to
\begin{equation}
\forall \z \in \C^N, \qquad (a,b) ~\odot~ \z = \Gamma^{-1}((a,b) ~\odot~ \Gamma(\z)) = \Gamma^{-1}(a r + b \mathbbm{1}) = a \z + b \u.
\end{equation}
This is well defined, because $a r + b \mathbbm{1}$ remains in $\Gamma(\C^N)$ for any interpolating curve $r$.

\subsection{A more general concept: configurations}

The two previous frameworks - landmarks configurations or interpolating curves - can be encompassed inside one common and more general framework.

\begin{definition}
	A \emph{configuration} of dimension $N \in \N^*$ in the plane is a complex vector $\z \in \C^N$.
\end{definition}

The notion of configuration plays a major role throughout this article and allows us to handle planar objects, such as curves or landmarks configurations, parameterized by $\{1,...,N\}$ through the configurations $\z \in \C^N$.

Set an integer $N \geq 1$. Let us consider a Hermitian matrix $\Phh = \Phh^* \in \C^{N \times N}$ that is positive-definite: ${\forall \z \in \C^N \setminus \{0\}},$ $\z^* \Phh \z > 0$. Note that $\Phh$ is not necessarily real-valued here. We endow $\C^N$ with the Hermitian inner product $\herPhi{\z}{\w} := \z^*\Phh \w$. Working with landmarks in the usual way, $\Phh$ refers to the identity matrix $\mathrm{Id} \in \C^{N \times N}$. In the case of interpolating curves, $\Phh$ is the Gram matrix of the basis functions $\phi_n$, assumed to be linearly independent, and to linearly generate $\mathbbm{1}$.

Now let $\u$ denote a non-zero configuration, that we call the \emph{shift configuration}, and consider the group action $\odot : (\mathcal{G},\circ) \times \C^N \to \C^N$ expressed as $(a,b) ~\odot~ \z \mapsto a \z + b \u,$ where $(\mathcal{G},\circ)$ is the group of similitudes. The shift configuration conveys an important geometrical interpretation, and is the vector determining the action of translations on $\C^N$, hence the term ``shift''. Together with $\Phh$, they will define the shape space built in next section. In our fundamental examples, we take $\u = (1,...,1)$ for landmarks, and $\u = \Gamma^{-1}(\mathbbm{1})$ for interpolating curves. In both cases, $\u$ specifies the coefficients of the unit vector or function in the corresponding basis of the representation.

In what follows, $\C\u$ denotes the complex vector line generated by $\u$, and $(\C \u)^{\perp_\Phh}$ the complex hyperplane of configurations orthogonal to $\u$.

\begin{definition}
	If $\z \in \C\u$, then $\z$ is said to be degenerate, and non-degenerate otherwise.
\end{definition}
In our examples of landmarks or interpolating curves, degenerate configurations represent objects that collapse to a single point in the plane, and are not interesting in practice. 

\begin{definition}
	\label{df:centre}
	A configuration $\z$ is said to be \emph{centered} if $\u^* \Phh \z = 0$ or, equivalently, $\z \in (\C \u)^{\perp_\Phh}$.
	
	\emph{Centering} $\z$ consists in orthogonally projecting $\z$ onto $(\C \u)^{\perp_\Phh}$.
\end{definition}
For any configuration $\z$, there exists a unique complex number $b \in \C$, called the \textit{center} of $\z$, so that $\z - b \u$ is orthogonal to $\u$. It is given by $b = \frac{\u^* \Phh \z}{|\u|^2_\Phh}$, and $\z - b \u$ is then the centered version of $\z$.
In the case of landmarks configurations, the center is the usual arithmetic mean. The operation $\z \mapsto \left(\z - \frac{ \sum\limits \z[i]}{n} \u\right)$ centers $\z$. Note that $|\u|_\Phh = \sqrt{n}$. Regarding interpolating curves, the center of $\z$ coincides with the temporal mean of $r$, given by
$\bar{r} := \int_0^1 r(t) \,\mathrm{d}t \in \C = \herH{\mathbbm{1}}{r} = \u^* \Phh \z = \frac{\u^* \Phh \z}{|\u|^2_\Phh}.$
Note that, contrarily to landmarks, $|\u|_\Phh = |\mathbbm{1}|_{\L^2([0,1],\C)} = 1$.

\begin{definition}
	A configuration $\z$ is called a \emph{pre-shape} if it is \emph{centered and normalized}
	$$\u^* \Phh \z = 0 \quad \text{and} \quad |\z|_\Phh = 1.$$
\end{definition}

To visualize the various operations, we encourage the reader to refer to Figure \ref{fig:preshaping}. We also summarize some notations in Table \ref{tab:notations}.

\section{Kendall's space of planar shapes}
\label{sec:kendall}

We now describe Kendall's shape space in the planar case \cite{dryden2016,srikla2016}. When developing this framework, we make a systematic use of the shift configuration $\u$ and Hermitian inner product $\Phh$ introduced in the previous section. In particular, interpolating curves can be referred to as configurations. We recall that we consider the action of similitudes written as $(a,b) ~\odot~ \z = a \z + b \u$.

\subsection{Pre-shape sphere and shape space}
A \textit{shape}, in the sense of Kendall, is an equivalence class of non-degenerate configurations considered up to similitudes (see Figure \ref{fig:notion_shape}). It is what remains after discarding redundant geometric information given by the position (center), scaling, and orientation. The shape space is then the quotient of the non-degenerate subspace by the action of similitudes.
We first get rid of translation and scaling, thus obtaining centered and normalized configurations which form the \textit{pre-shape sphere} (see Figure \ref{fig:preshaping}). Then, we quotient by the group of rotations
$U(1) := \{z \in \C ~|~ |z| = 1\}$
to obtain Kendall's shape space.

\begin{definition}
	The \emph{pre-shape sphere} $\S = \{\z \in \C^N ~|~ \u^* \Phh \z = 0, \z^* \Phh \z = 1 \}$ is the set of centered and normalized configurations $\z \in \C^N$. It is the quotient of $\C^N$ by the action of translations $\z \mapsto \z + b \u$, $b \in \C$, and of scalings $\z \mapsto \lambda ~ \z$, $\lambda \in \R^*$.
\end{definition}
Since it consists of the unit-norm elements of the complex hyperplane $(\C \u)^{\perp_\Phh}$ orthogonal to $\u$, $\S$ is a smooth compact real $(2N-3)$-manifold, as can be seen in
\begin{equation}
\S ~=~ \{|\z|_{\Phh} = 1 \} ~~ \cap ~~ (\C \u)^{\perp_\Phh}.
\end{equation}

Any non-degenerate configuration is uniquely associated to a \textit{pre-shape} by centering and then normalizing it (but the converse is not true).
\begin{definition}
The pre-shape uniquely associated to a non-degenerate configuration $\z \in \C^N$ is the projection of $\z$ onto $\S$, denoted by
\begin{equation}
\Pi_{\S}(\z) = \frac{\z_0}{|\z_0|}_\Phh, \qquad \z_0 = \z - \frac{\u^*\Phh \z}{|\u|_\Phh^2} \u,
\end{equation}
where $\z_0$ is the centered version of $\z$ (see Figure \ref{fig:preshaping} and Table \ref{tab:notations}).
\end{definition}

\begin{definition}
	The \emph{shape space}
	\begin{equation}
	\Sig := \S / U(1)
	\end{equation}
	is the pre-shape sphere quotiented by the action of the group of rotations $\z \mapsto \mathrm{e}^{\mathrm{i}\theta} ~ \z$, $\theta \in [0,2\pi)$.
	The orbit of a pre-shape $\z \in \S$ under the action of $U(1)$ is denoted by $[\z] \in \Sig$ and is called the \emph{shape} of $\z$. More generally, given a non-degenerate configuration $\z$, we define its shape $\shape{\z}$ to be that of the unique pre-shape associated to $\z$.
\end{definition}

Kendall's shape space $\Sig$ is a compact smooth real $(2N-4)$-manifold, identified to the complex projective space \cite{kendall1977,gallot2004,dryden2016,srikla2016}
\begin{equation}
\C\mathbb{P}^{N-2} = (\C^{N-1} \setminus \{0\}) / \C^* \simeq \mathbb{S}^{2N-3} / U(1).
\end{equation}

In this expression, $\C^{N-1} \setminus \{0\}$ refers to the set of centered non-degenerate configurations of landmarks. In the literature, a typical approach for quotienting out translations is to discard the last coordinate in $\z \in \C^N$. For centered configurations of landmarks, one indeed has that $\z[N-1] = -\sum_{n = 0}^{N-2} \z[n]$. In the general case, $\C^{N-1} \setminus \{0\}$ can be replaced by $(\C \u)^{\perp_\Phh} \setminus \{0\}$ and $\mathbb{S}^{2N-3}$ by $\S$, conserving the identification of $\Sig$ to $\C\mathbb{P}^{N-2}$.

\subsection{Distances in the shape space}
\label{sub:distances}
\begin{figure}[h]
	\centering
	\includegraphics[width=\linewidth]{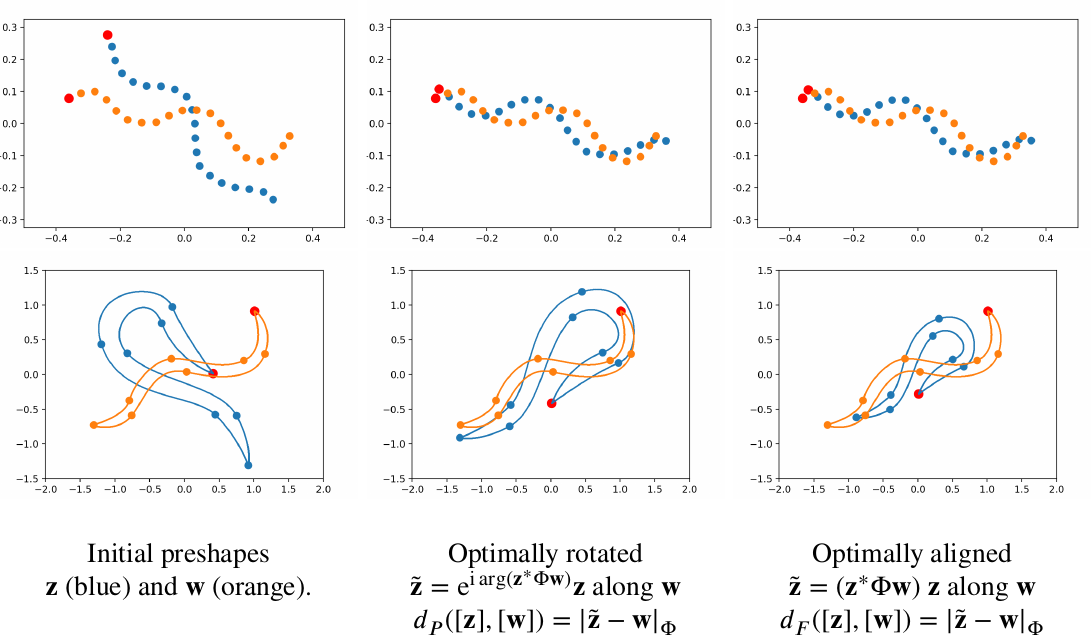}
	\caption{\small \textbf{Examples of optimal rotation and alignment.}  Upper row: landmarks with $N = 20$. Lower row: closed Hermite spline curves with $N = 16$ \cite{uhlmann2016}. Left column: original pre-shapes (centered and normalized) $\z$ in blue and $\w$ in orange. Middle column: optimal rotation of $\z$ along $\w$. Right column: optimal alignment of $\z$ along $\w$. The angles for both optimal rotation and alignment are the same and are equal to $\theta = \arg (\z^* \Phh \w)$. The optimal scaling in the alignment is the modulus $\lambda(\z,\w) = |\z^* \Phh \w|$; when $\z$ and $\w$ are highly correlated, $\lambda(\z,\w)$ is close to $1$ and the optimally aligned image of $\z$ is similar to the optimally rotated one.
		For landmarks (upper row), $|\z - \w|_\Phh = \sum_n |\z[n] - \w[n]|^2$ is the usual distance, here with $\z,\w$ being highly correlated.
		For Hermite-spline curves (lower row), the distance $|\z - \w|_\Phh$ is equal to the usual curve distance $|r - s|_{\L^2([0,1],\C)} = \sqrt{ \int_{0}^{1} |r(t)-s(t)|^2\,\mathrm{d}t } $ between their images $r,s$ through the isometry $\Gamma$ (Section \ref{subsub:splines}).}
	\label{fig:1}
\end{figure}

\begin{figure}[h]
	\centering
	\includegraphics{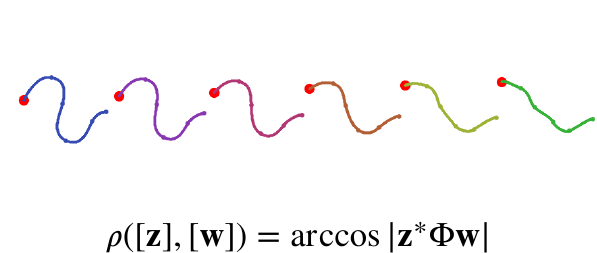}
	\caption{\small \textbf{Geodesic path on $\Sig$}. Pre-shapes corresponding to shapes regularly met along the geodesic path joining $[\z]$ to $[\w]$. The two endpoints are the pre-shapes $\tilde{\z} = \mathrm{e}^{ \mathrm{i} (\arg \z^* \Phh \w)} \z$ and $\w$ optimally rotated along each other. The pre-shapes themselves describe the shortest geodesic path that joins $\tilde{\z}$ to $\w$ on the pre-shape sphere $\S$. The distance $\rho([\z],[\w])$ then corresponds to the angle measured from the center of the pre-shape sphere along this geodesic arc. Here, we illustrate this concept with open Hermite-spline curves \cite{uhlmann2016,schoenberg1973}.}
	\label{fig:2}
\end{figure}

Three classical distances are usually defined on the shape space: the full (Procrustes), partial (Procrustes), and geodesic distances, denoted as $d_F$, $d_P$, and $\rho$, respectively \cite{dryden2016,srikla2016}. Their evaluations $\mathrm{dist}([\z],[\w])$ enjoy simple and closed-form expressions that involve only the pre-shapes $\z$ and $\w$. Readers interested in the practical use of these metrics and not in their geometrical definition may directly skip to \eqref{eq:d_P_expr}, \eqref{eq:d_F_expr}, and \eqref{eq:rho_expr}. For a quick geometrical intuition without mathematical details, we refer readers to Figures \ref{fig:1} and \ref{fig:2}.\newline

\paragraph{Distances and optimal transformations}
In this paragraph, $\z$ and $\w$ denote pre-shapes by default.

\begin{definition}[Partial distance and optimal rotation]
	Let $\z,\w \in \S$. We define
	\begin{equation} \label{eq:d_P}
	d_P([\z],[\w]) = \min \limits_{\theta \in [0,2\pi)} |\mathrm{e}^{ \mathrm{i}\theta} ~ \z - \w|_{\Phh}.
	\end{equation}
	We say that we optimally rotate $\z$ along $\w$ when, if unique, the \emph{optimal rotation angle} $\theta(\z,\w)$ is applied to $\z$ (see Figure \ref{fig:1}).
\end{definition}

\begin{definition}[Full distance and optimal alignment]
	\label{df:d_F}
	Let $\z,\w \in \S$. We define
	\begin{align}
	d_F([\z],[\w]) &= \min \limits_{a \in \C} |a ~ \z - \w|_{\Phh} \label{eq:d_F} \\
	&= |\mathrm{P}_{\C \z} \w - \w|_{\Phh} \label{eq:d_F_proj},
	\end{align}
	where $\mathrm{P}_{\C \z}$ is the orthogonal projection onto the complex vector line generated by $\z$ (with respect to $\Phh$).
	In \eqref{eq:d_F}, the \emph{optimal alignment factor} is given by $a(\z,\w) = \z^*\Phh \w$, and its modulus $\lambda(\z,\w)$ is called the \emph{optimal scaling factor} (see Figure \ref{fig:1}).
	We say that we \emph{optimally align $\z$ along $\w$} when $a(\z,\w)$ is applied to $\z$ to obtain $\mathrm{P}_{\C \z} \w = a(\z,\w) ~ \z = \z^* \Phh \w ~ \z$.
\end{definition}

\begin{proposition} \label{prop:link_d_P_d_F}
	Let $\z,\w \in \S$ and suppose that $\z^*\Phh\w \neq 0$.
	Then, the optimal rotation angle is given by
	\begin{equation}
\theta(\z,\w) = \arg (\z^*\Phh\w).
	\end{equation}
	As a consequence, the angle involved in the optimal alignment factor is the same as the optimal angle itself: $\arg(a(\z,\w)) = \theta(\z,\w)$.
	Otherwise, when $\z^*\Phh\w = 0$, the distances are maximal and equal to $d_F([\z],[\w]) = 1$ and $d_P([\z],[\w]) = \sqrt{2}$.
\end{proposition}
\begin{proof}
	If $\z^*\Phh\w = 0$, then we use the orthogonality of $\z$ and $\w$ to conclude. Otherwise, since
	\begin{equation}
\argmin \limits_{\theta \in [0,2\pi) } |\mathrm{e}^{ \mathrm{i}\theta} \cdot \z - \w|^2_{\Phh} = \argmax \limits_{\theta \in [0,2\pi)} \underset{\leq |\herPhi{\mathrm{e}^{ \mathrm{i}\theta} \cdot \z}{\w}|}{\underbrace{\Re \herPhi{\mathrm{e}^{ \mathrm{i}\theta} \cdot \z}{\w}}},
	\end{equation}
	$\theta$ is optimal when $\herPhi{\mathrm{e}^{ \mathrm{i}\theta} \cdot \z}{\w} = e^{-i\theta} \herPhi{\z}{\w} \in \R^+$; in other words, when $\theta = \arg \herPhi{\z}{\w}$.
\end{proof}

Note that the orthogonality condition $\z^*\Phh\w = 0$ means that $\z$ and $\w$ are decorrelated: whatever rotation we apply on $\z$, the distance $|\mathrm{e}^{ \mathrm{i}\theta} \cdot \z - \w|_\Phh$ remains maximal. In the real setting, this corresponds to the case where $\z$ is orthogonal to all rotations of $\w$ with respect to the real inner product (see Appendix \ref{sub:appendix_hermitian}).

\begin{corollary}
	\label{cor:d_P_d_F}
Let $\z,\w \in \S$. The distances have closed-form expressions given by
\begin{align}
d_P([\z],[\w]) &= \big|\frac{\z^*\Phh\w }{|\z^*\Phh\w|} \z - \w \big|_{\Phh} = \sqrt{2 - 2 |\z^*\Phh\w|} \in [0,\sqrt{2}], \label{eq:d_P_expr} \\
d_F([\z],[\w]) &= |(\z^*\Phh\w) ~ \z - \w |_{\Phh} = \sqrt{1 - |\z^*\Phh\w|^2} \in [0,1]. \label{eq:d_F_expr}
\end{align}
\end{corollary}

In Figure \ref{fig:1}, we illustrate the fact that the distances $d_P$ and $d_F$ measure the shortest reachable norm $|\tilde{\z} - \w|_\Phh$, if $\tilde{\z}$ is the image of $\z$ after any rotation, or scaling and rotation.

\begin{proposition}[Riemannian distance]
	\label{df:rho}
	Let $\z,\w \in \S$. The Riemannian distance in the shape space (see Figure \ref{fig:2}), also called geodesic distance and denoted by $\rho([\z],[\w])$, is equal to
	\begin{equation}
	\rho([\z],[\w]) = \arccos |\z^* \Phh \w| = \arccos \lambda(\z,\w) \quad \in [0,\pi/2]. \label{eq:rho_expr}
	\end{equation}
\end{proposition}

\begin{proposition}[Relationships between $d_F$, $d_P$, and $\rho$]
	The three distances $d_F$, $d_P$, and $\rho$ are related to each other as
	\begin{align}
	d_P^2 &= 2 - 2 \sqrt{1 - d_F^2}, \label{rel:d_P_d_F}\\
	d_F &= \sin \rho, \label{rel:d_F_rho}\\
	d_P &= 2 \sin (\rho/2). \label{rel:d_P_rho}
	\end{align}
\end{proposition}

We found it worth presenting the Riemannian distance $\rho$ alongside $d_F$ and $d_P$, although our 2DKSD method does not require the use of the Riemannian structure. In fact, the geometric structure of $\Sig$ is entirely inherited from that of $\S$: this justifies using the pre-shape sphere, whose geometry is well understood, to think about the shape space itself. In particular, the Riemannian structure of $\Sig$ is a quotient structure of $\S$, as discussed in Appendix \ref{sub:appendix_riem}. As a consequence, tangent spaces, geodesics, and exponentials benefit from explicit expressions on $\Sig$. For readers interested in the Riemannian structure of the shape space (not restricted to the planar case), we refer to \cite{dryden2016, srikla2016}, where the usual definition relying on landmarks is used. Examples of more advanced references about Riemannian geometry, in particular that of complex projective spaces, are \cite{gallot2004,gallier2011}.\newline

\section{Shape Dictionary}
\label{sec:shapedico}
The simplest dictionary learning approach on a dataset of configurations consists in considering them as elements of a point cloud in $(\C^N)^K \simeq (\R^{2N})^K$, and then applying an existing dictionary learning method. However, as discussed in the introduction, a pre-processing of the data can be necessary in order to attenuate differences of position, scaling, or orientation across the dataset. These geometric dissimilarities indeed lead to undesirable artefacts after learning, such as smoothed and unfaithful reconstructions or distorted and redundant atoms (in the sense that they share roughly the same shape but are associated to \textit{e.g.} different orientations).
An example of pre-processing is proposed in \cite{schmitter2018}, where the data are first optimally aligned
along a mean configuration $\z_\mathrm{mean}$, with $\shape{\z_\mathrm{mean}} \in \Sig$ being the Fréchet mean of $\shape{\z_1},...,\shape{\z_K} \in \Sig^K$ with respect to the distance $d_F$ (see Appendix \ref{sub:appendix_mean}).

A limitation of this kind of ``align-first'' approach is that the alignment step depends on the whole dataset, and needs to be started anew for each new data input. In Section \ref{sec:results}, we compare the results obtained by aligning first to those produced by our algorithm. Instead, we prefer to directly work in the shape space with suitable shape metrics, without having to pre-process the data. 

Our 2DKSD method takes advantage of \emph{complex weights} to rotate and scale the atoms before summing them to reconstruct each original shape. Hence, instead of pre-aligning the data, we rotate and scale \emph{inside the weighted sums $\D \alph_k$}, individually for each data shape. Also, we show that, for a good choice of the error metric that serves to compare the original and reconstructed shapes, \emph{the 2DKSD boils down to a nearly classical and very simple dictionary learning formulation.} Weights are however complex vectors, and the dataset and dictionary atoms must satisfy some (light) constraints imposed by Kendall's framework. This \textit{a priori} unexpected result, shown in Proposition \ref{prop:simple}, is interesting both mathematically and numerically.\newline

\paragraph{Statement of the Shape Dictionary}
We suppose that elements of the dataset are pre-shapes $\z_1,...,\z_K \in \S$ (centered and normalized). Let $\mathrm{dist}$ denote any distance on $\Sig,$ such as one of the three distances $d_F,d_P,\rho$.
\begin{definition}
The \emph{2DKSD general formulation} of the problem to be solved writes
\begin{equation} \label{eq:dico}
\inf \limits_{\D,\A : \substack{\d_j \in \S,\\\D \alph_k \in \S \\ |\alph_k|_0 \leq N_0}~~~~} \sum_{k=1}^K \mathrm{dist}\left([\z_k], [\D \alph_k]\right)^2.
\end{equation}
\end{definition}
The dictionary corresponds to $\D = (\d_1,...,\d_J) \in \C^{N \times J}$, whose atoms $\d_j \in \S$ are pre-shapes. Importantly, the weights $\alph_1,...,\alph_K \in \C^J$ are \emph{complex} numbers
which apply scalings and rotations to the atoms $\d_j$ before summing them to $\D \alph_k$.
This linear combination is sparse, since the (hard) sparsity constraint $|\alph_k|_0 \leq N_0$ enforces that at most $N_0$ coefficients are non-zero. We also impose $\D \alph_k$ and $\z_k$ to be pre-shapes before comparing their shapes. If $\hat{\D}$ and $\hat{\A}$ are (approximately) optimizers of \eqref{eq:dico}, the original shape $[\z_k]$ is reconstructed\footnote{When discussing the sparse coding step, we shall show that, for $\mathrm{dist} = d_F$, one can compute $\hat{\alph}_k$ so that the original pre-shape itself is reconstructed as $\hat{\D} \hat{\alph}_k$.} as $[\hat{\D} \hat{\alph}_k]$.

The unit-norm constraints imposed on $\d_j$ and $\D \alph_k$ are in fact nonessential. Up to a rescaling of the weights $\alph_k$, we can more generally consider $\d_j$ as a centered configuration of norm smaller than $1$, thus allowing possibly collapsing atoms $\d_j = \mathbf{0}$. This does not change the problem since this corresponds to a case where $\d_j$ would be ignored in the linear combinations of the original optimization problem \eqref{eq:dico}. Hence, we could replace $\d_j \in \S$ by $\d_j \perp_\Phh \u; |\d_j|_\Phh \leq 1$. We have decided instead to keep atoms as pre-shapes to have a uniform scaling across the dictionary.

Also, $\D \alph_k$ only needs to be a non-degenerate configuration in $\C^N \setminus \C \u$ for the shape $\shape{\D \alph_k}$ to be defined. Since each $\d_j$ is already centered, $\D \alph_k$ is necessarily centered, and so we only need to suppose that $\D \alph_k \neq \mathbf{0}$ does not collapse. We also note that $\Pi_\S(\D \alph_k) = \D \frac{\alph_k}{|\D \alph_k|}_\Phh$ is then simply the normalized version of $\D \alph_k$. Hence, the infimum in \eqref{eq:dico} is equal to
\begin{equation} \label{eq:dico2}
\inf \limits_{\D,\A : \substack{\d_j \in \S\\\D \alph_k \in \S \\ |\alph_k|_0 \leq N_0}~~~~} \sum_{k=1}^K \mathrm{dist}\left([\z_k], [\D \alph_k]\right)^2 = \inf \limits_{\D,\A : \substack{\d_j \in \S\\\D \alph_k \neq \mathbf{0} \\ |\alph_k|_0 \leq N_0}~~~~} \sum_{k=1}^K \mathrm{dist}\left([\z_k], [\D \frac{\alph_k}{|\D \alph_k|}_\Phh] \right)^2.
\end{equation}

\begin{proposition}\label{prop:simple}
Using the full Procrustes distance $\mathrm{dist} = d_F$ as minimization criterion, \eqref{eq:dico2} leads to the \emph{2DKSD simple formulation}
\begin{equation} \label{eq:dico_simple}
\inf \limits_{\D : \d_j \in \S} \sum_{k=1}^K \min \limits_{\alph_k : |\alph_k|_0 \leq N_0 } \left|\z_k - \D \alph_k \right|_\Phh^2,
\end{equation}
where we recall that $\z_k$ are pre-shapes. This formulation is equivalent to the one in \eqref{eq:dico_simple0}.
\end{proposition}
\begin{proof}
We express the minimization problem \eqref{eq:dico2} as
\begin{equation}
\inf \limits_{\D : \left\{ \substack{\d_j \perp_\Phh \u \\ |\d_j|_\Phh \leq 1} \right. } \sum_{k=1}^K \inf \limits_{\alph_k : \left\{ \substack{\D \alph_k \neq \mathbf{0} \\ |\alph_k|_0 \leq N_0} \right.} d_F\left([\z_k], [\D \frac{\alph_k}{|\D \alph_k|}_\Phh] \right)^2
\end{equation}
since, after fixing $\D$, the minimization breaks into $K$ independent elementary terms that correspond to a sparse coding of the shapes $[\z_k]$. Each term has the form
\begin{equation} \label{eq:dico_elementary}
\inf \limits_{\alph : \left\{ \substack{\D \alph \neq \mathbf{0} \\ |\alph|_0 \leq N_0} \right.} d_F \left([\z], [\D \frac{\alph}{|\D \alph|}_\Phh]  \right)^2 = \inf \limits_{\alph : \left\{ \substack{\D \alph \neq \mathbf{0} \\ |\alph|_0 \leq N_0} \right.} |\z - \mathrm{P}_{\C \D \alph} \z|^2_\Phh = \inf \limits_{\alph : |\alph|_0 \leq N_0} |\z - \mathrm{P}_{\C \D \alph}\z|^2_\Phh,
\end{equation}
where we successively applied \eqref{eq:d_F_proj}, used that the vector line generated by $\D \frac{\alph}{|\D \alph|}_\Phh$ and $\D \alph$ is the same, and dropped\footnote{The case $\D \alph = \mathbf{0}$ does not induce a lower infimum value because $|\z - \mathrm{P}_{\C \D \alph} \z|^2_\Phh = |\z|^2_\Phh$ is then maximal.} the inequality $\D \alph \neq \mathbf{0}$.
Then, we apply the results of Lemma \ref{lm:2_pbs} and find that
\begin{align}
\min \limits_{\alph : \left\{ \substack{\D \alph \neq \mathbf{0} \\ |\alph|_0 \leq N_0} \right.} d_F \left([\z], [\D \frac{\alph}{|\D \alph|}_\Phh]  \right)^2 & = \min \limits_{\alph : \left\{ \substack{\D \alph \neq \mathbf{0} \\ |\alph|_0 \leq N_0} \right.} |\z - \mathrm{P}_{\C \D \alph} \z|^2_\Phh = \min \limits_{\alph : |\alph|_0 \leq N_0} |\z - \mathrm{P}_{\C \D \alph}\z|^2_\Phh \\
& = \min \limits_{|\alph|_0 \leq N_0} |\z - \D \alph|^2_\Phh,
\end{align} where the three first terms share a common minimizer. Therefore, a new formulation of the original problem is given by \eqref{eq:dico_simple} and, equivalently, by \eqref{eq:dico_simple0}.
~~\newline
\end{proof}

\section{Dictionary update and sparse coding}
\label{sec:algorithms}

We numerically solved the 2DKSD problem \eqref{eq:dico_simple} involving the full distance $d_F$. Our code is fully available online.

As summarized in Algorithms \ref{algo:2DKSD} and \ref{algo:ORMP}, the implementation is directly adapted from the algorithm used in the \verb|SPAMS| software \cite{mairal2009,mairal2014} to a complex Hermitian framework. Recall that $\D = (\d_1,...,\d_J)$ is the dictionary and $\A = (\alph_1,...,\alph_K)$ contains the weighting complex vectors.
To minimize the corresponding loss functional
\begin{equation}
\label{eq:loss}
E(\D,\A ) = \sum \limits_{k = 1}^K |\z_k - \D \alph_k|_\Phh^2,
\end{equation}
we use a classical procedure. After initializing the dictionary with random elements of the dataset, the algorithm alternates between a sparse coding step in which $\A$ minimizes \eqref{eq:loss},
and where $\D$ is fixed \cite{lee2006,mairal2009,mairal2014}, and a dictionary update step where $\A$ is fixed.

\subsection{Dictionary update with MOD}
To update the dictionary, it is natural to think about performing gradient descent over $\D$ for updating the dictionary. Let us denote by $\Zbf = (\z_1,...,\z_K) \in \C^{N \times K}$ the matrix whose columns contain the dataset. The gradient can be conveniently explicited as
$\mbox{\boldmath{$\nabla$}}_\D E(\D,\A) = 2 \Phh \left(\D \A \A^* - \Zbf \A^* \right).$
Here, however, we prefer to rely on the Method of Optimal Directions (MOD) introduced by \cite{engan1999}, as done in the \verb|SPAMS| software \cite{mairal2009,mairal2014}. Given a fixed value of $\A$, the dictionary is updated so that it solves $\min_\D E(\D,\A)$, and its columns are then projected back onto $\S$. It is known that, in the landmarks case $\Phh = \mathrm{Id}$, a particular solution to the least-squares problem $\min \limits_{\D \in \C^{N \times J}} E(\D,\A)$ is
\begin{equation}
\label{eq:dico_update}
\hat \D = \Zbf \A^{+} = \Zbf \Vbf \Sigma^{+} \Ubf^*,
\end{equation}
where $\A^{+} = \Vbf \Sigma^{+} \Ubf^*$ is the pseudo-inverse of $\A$, and the SVD decomposition of $\A$ is $\A = \Ubf \Sigma \Vbf^*$. The pseudo-inverse $\Sigma^+$ corresponds to the matrix $\Sigma$ where non-zero diagonal elements are replaced by their multiplicative inverse.
We assert in Lemma \ref{lm:dico_update} of Appendix \ref{appendix:two_lemmas} that, interestingly, when $\Phh \neq \mathrm{Id}$, the update is the same and does not depend on $\Phh$.

Let us remark that the solution \eqref{eq:dico_update} cancels the gradient (one can see it by checking that $\A^+ \A \A^* = \A^*$), and also that the update does not involve any previous value assigned to $\D$. When $(\A \A^*)$ is invertible\footnote{This happens whenever the rank of $\A$ is equal to its number of lines or, equivalently, when the columns of $\A^*$ are independent.}, we have $\A^{+} = \A^* (\A \A^*)^{-1}$ and the solution is then unique and equal to $\hat \D= \Zbf \A^* (\A \A^*)^{-1}$.

After this first operation, the non-zero columns of the dictionary are projected back onto the pre-shape sphere by applying the normalization
$$\forall j = 1,...,J, \quad \d_j \leftarrow \d_j / |\d_j|_\Phh.$$
This is sufficient as \eqref{eq:dico_update} already centers the atoms, since the data themselves are centered. We replace columns which have become zero, as well as under-utilized ones, by pre-shapes randomly picked in the original dataset.

\begin{algorithm}
	\caption{2D Kendall Shape Dictionary with Method of Optimal Directions}
	\label{algo:2DKSD}
	\begin{algorithmic}[1]
		\Procedure{2DKSD with MOD}{}\newline
		\textbf{Dataset:} pre-shapes $\z_1,...,\z_K$ \newline
		\textbf{Parameters:} number of iterations $T$, sparsity $N_0$, number of atoms $J$ \newline
		\textbf{Initialization:} initial dictionary $\D_0$ of pre-shapes \newline
		\textbf{Output:} dictionary $\D = (\d_1,...,\d_J)$ of pre-shapes \newline
		 
		\State{$\D \leftarrow \D_0$}
			\Comment{Dictionary initialization}
		\For{$t = 1,...,T$}			
			\For{$k = 1,...,K$}
				\State{$\alph_k \simeq \argmin \limits_{|\alph|_0 \leq N_0 } |\z_k - \D \alph|_\Phh^2 $}
					\Comment{Sparse coding with {\textsc{complex}} ORMP (Algo. \ref{algo:ORMP})}
			\EndFor
			\State{$\D \leftarrow \Pi_{\S} \left(\Zbf \A^{+} \right)$, where $\A = (\alph_1,...,\alph_K)$.}
				\Comment{Dictionary update with MOD (Eq. \eqref{eq:dico_update})}
		\EndFor	

		\EndProcedure
	\end{algorithmic}
\end{algorithm}

\subsection{Sparse coding with ORMP}
\label{sub:sparse_coding}
The sparse coding problem for a data point $\z$ is formulated as
\begin{equation}
\label{eq:sparse_coding}
\min \limits_{\substack{\alph \in \C^J \\ |\alph|_0 \leq N_0}} |\z - \D \alph|^2_\Phh,
\end{equation}
where $\D$ is known. Equivalently, the non-zero coefficients of the solution $\hat \alph$ correspond to a set of indices $\hat{I}$ that solves
\begin{equation}
\label{eq:sparse_coding_I}
\min \limits_{\substack{I \subset \{1,...,J\} \\ |I| \leq N_0}} |\z - \mathrm{P}_{\C \{\d_j\}_{j \in I}}(\z)|^2,
\end{equation}
where $\C \{\d_j\}_{j \in I}$ denotes the complex vector space generated by the columns indexed by $I$.

Let us suppose that a solution $\hat{\alph}$ has been found. Then, the shape $[\z]$ is approximately reconstructed as $[\D \frac{\hat{\alph}}{|\D \hat{\alph}|}_\Phh]$, whenever $\D \hat{\alph} \neq \mathbf{0}$. Note that, in fact, the original pre-shape $\z$ itself (with preserved orientation) can also be approximated as $\D \frac{\hat{\alph}}{|\hat{\D} \hat{\alph}|}_\Phh$: there is no need to rotate $\D \hat{\alph}$ in order to bring it close to the original pre-shape. As stated in the proof of Lemma \ref{lm:2_pbs}, this is because the solution $\hat{\alph}$ satisfies $\mathrm{P}_{\C \D \hat{\alph}}\z = \D \hat{\alph}$, implying that $\D \hat{\alph}$ corresponds to the optimal alignment of the pre-shape $\D \frac{\hat{\alph}}{|\D \hat{\alph}|}_\Phh$ along $\z$ (see Definition \ref{df:d_F}). Also, $\D \frac{\hat{\alph}}{|\D \hat{\alph}|}_\Phh$ is optimally rotated along $\z$ (see Proposition \ref{prop:link_d_P_d_F}).\newline

\paragraph{Order Recursive Matching Pursuit}
A possible way to approximate the solution of \eqref{eq:sparse_coding} is to rely on a complex version of Orthogonal Matching Pursuit (OMP) \cite{pati1993,peyre2011} adapted to the Hermitian framework $(\C^n,\Phh)$. Instead, we prefer to use Order Recursive Matching Pursuit (ORMP), with a Cholesky-based optimisation \cite{cotter1999,mairal2014}, as it was empirically found to be more efficient than an adapted complex OMP. The ORMP algorithm is explained in \cite{cotter1999} and a (non-optimized) implementation is presented in Algorithm \ref{algo:ORMP}. ORMP starts with an empty set of indices $I[0] = \emptyset$, a remainder $\rbf^{(0)} = \z$, and a family of vectors $\{\de_{1}^{(0)},...,\de_{J}^{(0)}\} = \{\d_{1},...,\d_{J}\}$. At each step, the algorithm increases $I[l-1]$ by a new element $j[l]$ maximizing $\tfrac{|\de_j^{(l-1) *} \Phh \rbf^{(l-1)} |}{|\de_j^{(l-1)}|_\Phh}$ (wherever defined), in such a manner that
$I[l] = I[l-1] \cup \{j[l]\}$ minimizes \eqref{eq:sparse_coding_I} under the nesting constraint $I[l-1] \subsetneq I$ and $|I[l]| = l$.
Then, it updates the vectors $\{\de_{1}^{(l)},...,\de_{J}^{(l)}\}$ in a way that is similar to the Gram-Schmidt process\footnote{The difference being that the order is imposed by $j[1],...,j[N_0]$, so that the final orthonormal family is $\{\de_{j[1]}^{(0)},...,\de_{j[N_0]}^{(N_0-1)}\}$.}.
It stops either when $N_0$ indices have been found, or when $\z$ is spanned by $l_{\mathrm{max}}$ independent atoms and a new index $j[l_{\mathrm{max}} + 1]$ does not contribute in minimizing \eqref{eq:sparse_coding_I} further. Finally, the output weight $\alph$ is solution to $\mathrm{P}_{\C \{\d_j\}_{j \in I[l_{\mathrm{max}}]}}(\z) = \D_{I[l_{\mathrm{max}}]} \alph_{I[l_{\mathrm{max}}]}$. Its non-zero coefficients are given by
\begin{equation}
\label{eq:alpha}
\alph_{I[l_{\mathrm{max}}]} = (\D_{I[l_{\mathrm{max}}]}^* \Phh \D_{I[l_{\mathrm{max}}]})^{-1} \D_{I[l_{\mathrm{max}}]}^* \Phh ~ \z.
\end{equation}

\paragraph{Optimizing ORMP}
Thanks to the so-called Cholesky optimisation of ORMP \cite{cotter1999,mairal2009,mairal2014}, it is in fact possible to speed up computations considerably. The Gram matrix $\D^* \Phh \D$ is pre-computed before coding the data $\z$ (and re-used for other data). At each step $1 \leq l \leq N_0$, only the elements
$$\frac{\de_{j[l]}^{(l-1) *} \Phh \d_j }{|\de_{j[l]}^{(l-1)}|_\Phh}
= \frac{\de_{j[l]}^{(l-1) *} \Phh \de_j^{(l-1)}}{|\de_{j[l]}^{(l-1)}|_\Phh},
\qquad 
|\de_j^{(l)} |_\Phh^2,
\qquad
\d_j^* \Phh \rbf^{(l-1)},
\quad \text{and} \quad
\frac{\de_j^{(l-1) *} \Phh \rbf^{(l-1)} }{|\de_j^{(l-1)}|_\Phh}
= \frac{\d_j^* \Phh \rbf^{(l-1)} }{|\de_{j}^{(l-1)}|_\Phh}
$$
are updated, for $j = 1,...,J$.
The coefficients of the decomposition of $\qbf^{(l)} = \frac{\de_{j[l]}^{(l-1)}}{|\de_{j[l]}^{(l-1)}|_\Phh}$ in the linearly independent family $\{\d_{j[1]},..., \d_{j[l]}\}$ are also iteratively computed. If these coefficients are stacked into a column vector $\cbf^{(l)} \in \C^{N_0}$ where the last $N_0 - l$ coefficients are zero, the matrix $\mathbf{U} = (\cbf^{(1)},...,\cbf^{(N_0)}) \in \C^{N_0\times N_0}$ is upper triangular with strictly positive real scalars on the diagonal. Furthermore, it is the inverse conjugate of the Cholesky factor $\mathbf{L}$ involved in the factorisation $\D_{I[N_0]}^* \Phh \D_{I[N_0]} = (\mathbf{U}^*)^{-1} \mathbf{U}^{-1} = \mathbf{L} \mathbf{L}^*$. The final weight $\alph$ is then given by
$$\alph_{I[N_0]} = \mathbf{U} ~ \left[\frac{\de_{j[1]}^{(0) *} \Phh \rbf^{(0)} }{|\de_{j[1]}^{(0)}|_\Phh},...,\frac{\de_{j[N_0]}^{(N_0-1) *} \Phh \rbf^{(N_0-1)} }{|\de_{j[N_0]}^{(N_0-1)}|_\Phh} \right]^T.$$
All computations can be inferred from the updates in Algorithm \ref{algo:ORMP} and are not detailed here. They can be transparently investigated in the available source code.

\begin{algorithm}
	\caption{Complex Order Recursive Matching Pursuit}
	\label{algo:ORMP}
	\begin{algorithmic}[1]
		\Procedure{Complex ORMP}{}\newline
		\textbf{Data:} one pre-shape $\z$ \newline
		\textbf{Parameters:} sparsity $N_0$ \newline
		\textbf{Input dictionary:} $\D = (\d_1,...,\d_J)$ \newline
		\textbf{Output:} weight $\alph \in \C^J$ \newline

		\State{$l = 1,\quad l_{\mathrm{max}} = N_0, \quad I[0] = \emptyset$}
		\State{$\rbf^{(0)} = \z, \quad \de_j^{(0)} = \d_j$ for $j = 1,...,J$}
			\Comment{Initialization}
		
		\While{$l \leq N_0$}
			\State{$j[l] =
				\left\{ \argmax \tfrac{|\de_j^{(l-1) *} \Phh \rbf^{(l-1)} |}{|\de_j^{(l-1)}|_\Phh} ~\vert~ j \notin I[l-1], |\de_{j}^{(l-1)}|_\Phh \neq 0 \right\}
				$}
			\State{$I[l] = I[l-1] \cup \{j[l]\}$}
			\If{this maximum is zero}
				\State{$l_{\mathrm{max}} = l-1$}
				\State{\textbf{break}}
			\EndIf
			\State{$\qbf^{(l)} = \frac{\de_{j[l]}^{(l-1)}}{|\de_{j[l]}^{(l-1)}|_\Phh}$}
				\Comment{New orthonormal basis element}
			\State{$\de_j^{(l)} = \de_j^{(l-1)} - (\qbf^{(l)*}~\Phh~\de_j^{(l-1)}) ~ \qbf^{(l)} \quad \text{for } j = 1,...,J$}
				\Comment{Project on the orthogonal subspace}
			\State{$\rbf^{(l)} = \rbf^{(l-1)} - (\qbf^{(l)*}~\Phh~ \rbf^{(l-1)}) ~ \qbf^{(l)} \quad \text{for } j = 1,...,J$}
				\Comment{Update the remainder}
			\State{$l \leftarrow l+1$}
		\EndWhile
		\State{Use \eqref{eq:alpha} to build $\alph$}
			\Comment{Solve $\mathrm{P}_{\C \{\d_j\}_{j \in I[l_{\mathrm{max}}]}}(\z) = \D_{I[l_{\mathrm{max}}]} \alph_{I[l_{\mathrm{max}}]}$}
		\EndProcedure
	\end{algorithmic}
\end{algorithm}

\section{Computational results}
\label{sec:results}

We ran our algorithm on $5$ datasets that vary in size, type and shape.\footnote{Experimental results and datasets from this section can be found at https://github.com/ansonang3/2DKSD.}

\begin{enumerate}
\item[--] Dataset 1\footnote{Images were extracted from videos available from the \textit{C. elegans} behavioral database \cite{yemini2013}.} consists of $K > 5500$ skeletons of \textit{Caenorhabditis elegans} nematodes delineated as discretely-defined configurations of $N = 20$ landmarks. The dataset results from the concatenation of videos featuring freely crawling nematodes aged between $2$ and $18$ days and presenting various locomotion behaviors. The dataset was reconstructed using $N_0 = 5$ atoms out of the $J = 10$ atoms learned by 2DKSD.
\item[--] Dataset 2\footnote{Data courtesy of \cite{migliozzi2018}.} contains $K > 6300$ skeletons of $4$ \textit{Caenorhabditis elegans} nematodes delineated as continuously-defined open Hermite-spline curves \cite{uhlmann2016,schoenberg1973} with $N = 12$ degrees of freedom. The animals lied in a shared container and were constrained by the lack of space, leading to looping or wavy shapes. The dataset was reconstructed using $N_0 = 5$ atoms out of the $J = 10$ atoms learned by 2DKSD.
\item[--] Dataset 3\footnote{Data courtesy of \cite{stegmann2002}.} features $K = 40$ hands of $4$ different people, as outlined by $N = 56$ landmarks. The dataset was reconstructed using $N_0 = 3$ atoms out of the $J = 10$ atoms learned by 2DKSD.
\item[--] Dataset 4 and 4b\footnote{Data extracted from the binary images of the Kaggle leaf dataset https://www.kaggle.com/c/leaf-classification.} are constituted of $K > 1500$ leaves outlines, and come in the form of configurations of $N = 200$ landmarks and cubic B-splines \cite{unser1993,brigger2000} with $N = 40$ degrees of freedom, respectively. For most shapes, the first landmark either marks the stem or the leaf tip. Both datasets were (independently) reconstructed using $N_0 = 5$ atoms out of the $J = 20$ atoms learned by 2DKSD.
\item[--] Dataset 5 is a synthetic dataset generated by deforming original silhouettes extracted from the binary images of the MPEG-7 database\footnote{Data courtesy of \cite{latecki2000}. See http://www.dabi.temple.edu/~shape/MPEG7/dataset.html}. For each original pre-shaped configuration $\z_\mathrm{or} = (\z_\mathrm{or}^1,...,\z_\mathrm{or}^N) \in \C^N$, $N$ random Gaussian planar vectors were generated and smoothed with the kernel $K = (|\z_\mathrm{or}^i - \z_\mathrm{or}^j|^2)_{i,j}$ of the squared distances between landmarks. These deformations were then added to $\z_\mathrm{or}$. The process was repeated $10$ times, resulting in a dataset of $K = 14000$ shapes featuring $70$ classes of objects, each containing $200$ configurations of $N = 200$ landmarks. The dataset was reconstructed using $N_0 = 30$ atoms out of the $J = 230$ atoms learned by 2DKSD.\newline
\end{enumerate}

\paragraph{Comparison to the ``align-first'' method }

The problem, that we call ``align-first'', relating most to the one we are solving, is stated as
\begin{equation}
\label{eq:align_first}
\inf \limits_{ \D', \A' : \substack{\hskip-.5cm \d'_j \in \S \\ |\alph'_k|_0 \leq N_0 } } \sum_{k=1}^K \left|\widetilde{\z}_k - \D' \alph'_k \right|_\Phh^2, \qquad \text{where } \alph'_k \in \R^{J},
\end{equation}
where $\A' = (\alph'_1,...,\alph'_K)$, and $\widetilde{\z}_1,...,\widetilde{\z}_K \in \S$ are pre-shapes optimally rotated along a reference pre-shape. Typically, we chose to rotate them along a mean pre-shape $\z_\mathrm{mean}$ (see Appendix \ref{sub:appendix_mean}).
This standard problem is essentially the same as in \eqref{eq:dico_simple}, except that the original dataset is pre-processed and, more notably, that the weights $\alph'_k$ are real vectors.

In practice, we took advantage of the MOD combined with the real version of the Cholesky-optimised ORMP provided by the \verb|SPAMS| toolbox \cite{mairal2009}. This is an especially appropriate comparison, since our algorithm is a direct adaptation of the latter to the Hermitian framework. Thanks to an isometric transformation through $\sqrt{\Phh}$, \eqref{eq:align_first} could be reformulated as a standard $\ell^2$ problem.\newline

\paragraph{Comparison to complex PCA}
As a valuable and complementary comparison, we also considered the complex version of principal component analysis. It also relies on complex weights, but atoms are chosen in a specific way, that we recall here in our setting. Suppose that the average configuration has been discarded in each element of the dataset, and let us still use $\z_1,...,\z_K$ to denote this new dataset (the average configuration coordinates are simply the coefficient-by-coefficient arithmetic means). We find the first complex PCA mode by solving
\begin{equation}
\label{eq:complex_PCA}
\argmin \limits_{|\w_1|_\Phh \leq 1} \sum_{k = 1}^{K} |\z_k - (\w_1^* \Phh \z_k) \w_1|_\Phh^2 = \argmin \limits_{|\w_1|_\Phh \leq 1} \sum_{k = 1}^{K} |\z_k|_\Phh^2 - |\w_1^* \Phh \z_k|^2 = \argmax \limits_{|\w_1|_\Phh \leq 1} \sum_{k = 1}^{K} |\w_1^* \Phh \z_k|^2.
\end{equation}
Let us set $\Zbf = (\z_1,...,\z_K)$ as before. The matrix $\Zbf \Zbf^* \Phh$ is self-adjoint with respect to the Hermitian product $\Phh$, because $\forall \x,\y \in \C^N, \herPhi{\Zbf \Zbf^* \Phh ~ \x }{\y} = \herPhi{\x}{\Zbf \Zbf^* \Phh ~ \y}$. By the Spectral Theorem, this implies that there exists a $\Phh$-orthonormal basis $\Vbf \in \C^{N \times N}$ with $\Vbf^* \Phh \Vbf = \mathrm{Id}$ in which $\Zbf \Zbf^* \Phh$ is expressed as a diagonal matrix $\Dbf$:
$$\Zbf \Zbf^* \Phh = \Vbf \Dbf \Vbf^* \Phh = \Vbf \Dbf \Vbf^{-1}.$$
Since $\Zbf \Zbf^* \Phh$ is positive semi-definite with respect to $\Phh$, we know that $\Dbf = \mathrm{diag}(\lambda_1,...,\lambda_N)$ has (real) non-negative values that can be ordered as $\lambda_1 \geq ... \geq \lambda_N \geq 0$.
A solution of \eqref{eq:complex_PCA} then corresponds to a unit-norm eigenvector of $\Zbf \Zbf^* \Phh$ corresponding to $\lambda_1$.
The second complex PCA mode $\w_2$, is the solution to a problem analogous to \eqref{eq:complex_PCA}, except that we add the constraint $\w_2 \perp_\Phh \w_1$. It is given by a unit-norm eigenvector corresponding to $\lambda_2$. Further modes are iteratively found with the constraint to be $\Phh$-orthogonal to previous modes.

The dictionary is then the collection of the modes $(\w_1,...,\w_{J})$ and the best $N_0$-term approximation of $\z_k$ is
$\sum_{j = 1}^{N_0} (\w_j^* \Phh \z_k) \w_j$. The corresponding loss is then equal to $\sum_{k = 1}^{K} |\z_k|_\Phh^2 - \lambda_1 - ... - \lambda_{N_0}$. As in the previous paragraph, we employ $\sqrt{\Phh}$ to transform the problem into a standard form suitable for computations.

\subsection{Results}
\begin{figure}[H]
	\centering
	\includegraphics[clip,trim = {0cm .8cm 0cm 0cm}, width=\linewidth]{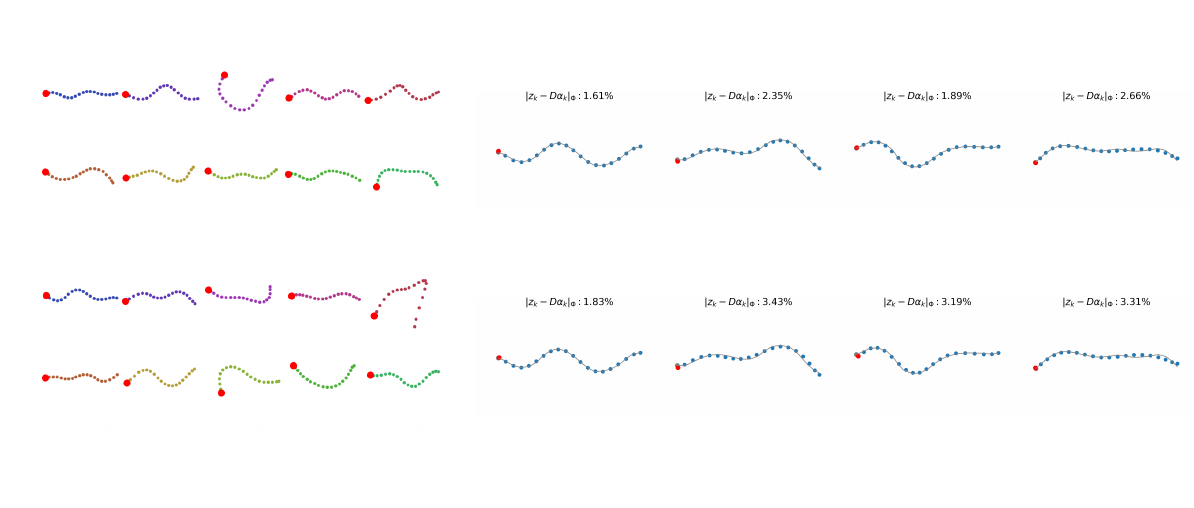}
	\caption{\small\textbf{Dataset 1.} Up: our method. Down: align-first method. Left: shape dictionary of $J = 10$ atoms, taking $N_0 = 5$ out of them to reconstruct the $K > 5500$ shapes represented as configurations of dimension $N = 20$. Right: four examples of reconstruction (blue) over the original data (gray). The RMSEs are $2.36\%$ and $4.19\%$, respectively.}
	\label{fig:worms}
\end{figure}

\begin{figure}[H]
	\centering
	\includegraphics[clip,trim = {0cm .8cm 0cm 0cm}, width=\linewidth]{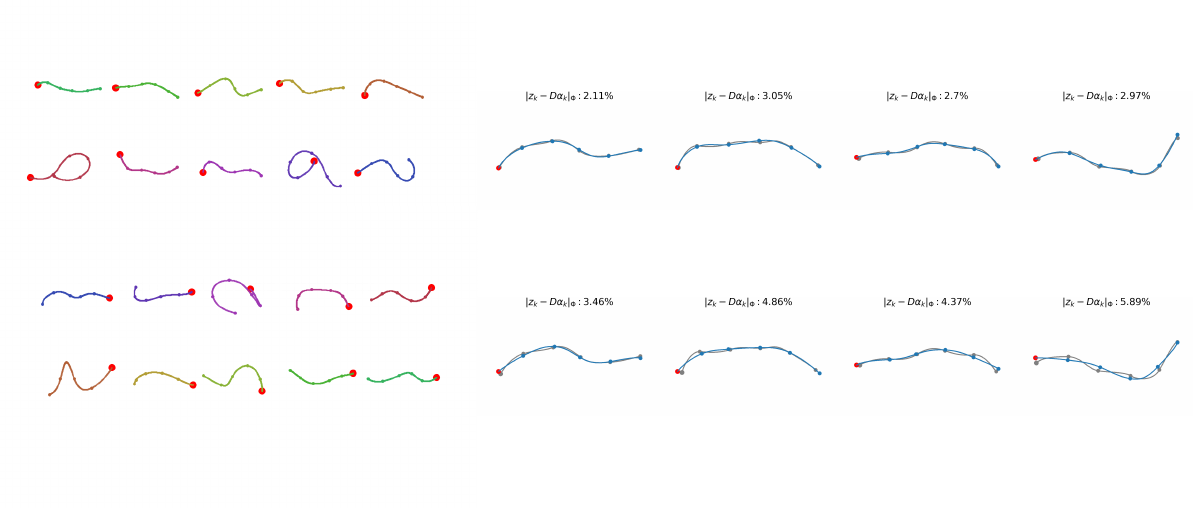}
	\caption{\small\textbf{Dataset 2.} Up: our method. Down: align-first method. Left: shape dictionary of $J = 10$ atoms, taking $N_0 = 5$ out of them to reconstruct the $K > 6300 $ shapes represented as configurations of dimension $N = 12$. Right: four examples of reconstruction (blue) over the original data (gray). The RMSEs are $4.04 \%$ and $7.95 \%$, respectively. We display the curves and the control points.}
	\label{fig:6376_open_worms}
\end{figure}

\begin{figure}[H]
	\centering
	\includegraphics[clip,trim = {0cm .8cm 0cm 0cm}, width=\linewidth]{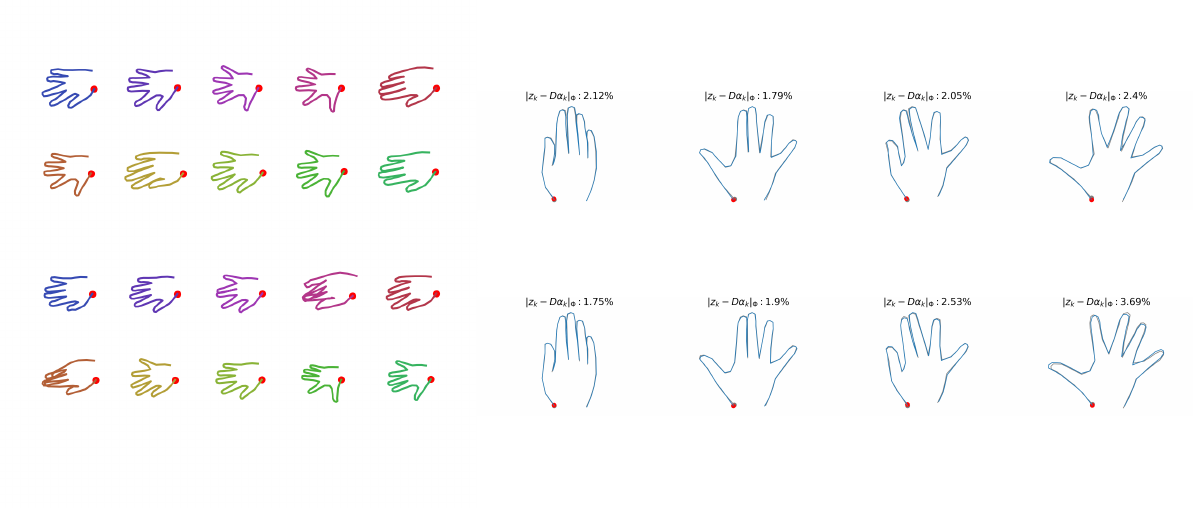}
	\caption{\small\textbf{Dataset 3.} Up: our method. Down: align-first method. Left: shape dictionary of $J = 10$ atoms, taking $N_0 = 3$ out of them to reconstruct the $K = 40$ shapes represented as configurations of dimension $N = 56$. Right: four examples of reconstruction (blue) over the original data (gray). The RMSEs are $2.48\%$ and $2.73\%$, respectively. We display configurations of landmarks as continuous curves.}
	\label{fig:hands}
\end{figure}

\begin{figure}[H]
	\centering
	\includegraphics[width=\linewidth]{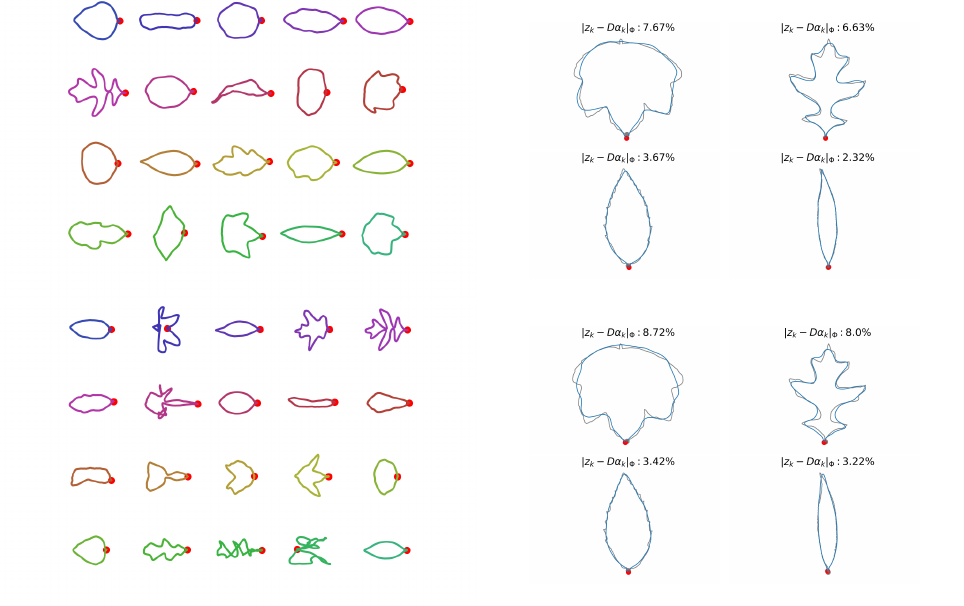}
	\caption{\small\textbf{Dataset 4.} Up: our method. Down: align-first method. Left: shape dictionary of $J = 20$ atoms, taking $N_0 = 5$ out of them to reconstruct the $K > 1500$ shapes represented as configurations of dimension $N = 200$. Right: four examples of reconstruction (blue) over the original data (gray). The RMSEs are $6.45\%$ and $7.12\%$, respectively. We display the configurations of landmarks as continuous curves.}
	\label{fig:leaves_sym}
\end{figure}

\begin{figure}[H]
	\centering
	\includegraphics[width=\linewidth]{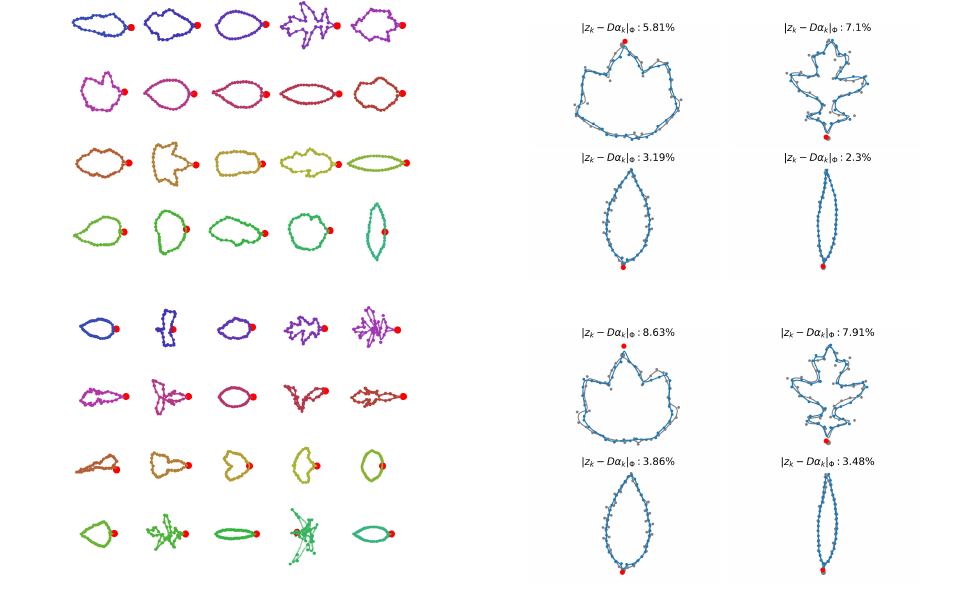}
	\caption{\small\textbf{Dataset 4b.} Up: our method. Down: align-first method. Left: shape dictionary of $J = 20$ atoms, taking $N_0 = 5$ out of them to reconstruct the $K > 1500$ shapes represented as configurations of dimension $N = 40$. Right: four examples of reconstruction (blue) over the original data (gray). The RMSEs are $6.41\%$ and $7.01\%$, respectively. We display the curves and also the control points, which explains the non-smooth effects.}
	\label{fig:leaves_Bsplines}
\end{figure}

\begin{figure}[H]
	\centering
	\includegraphics[width=\linewidth]{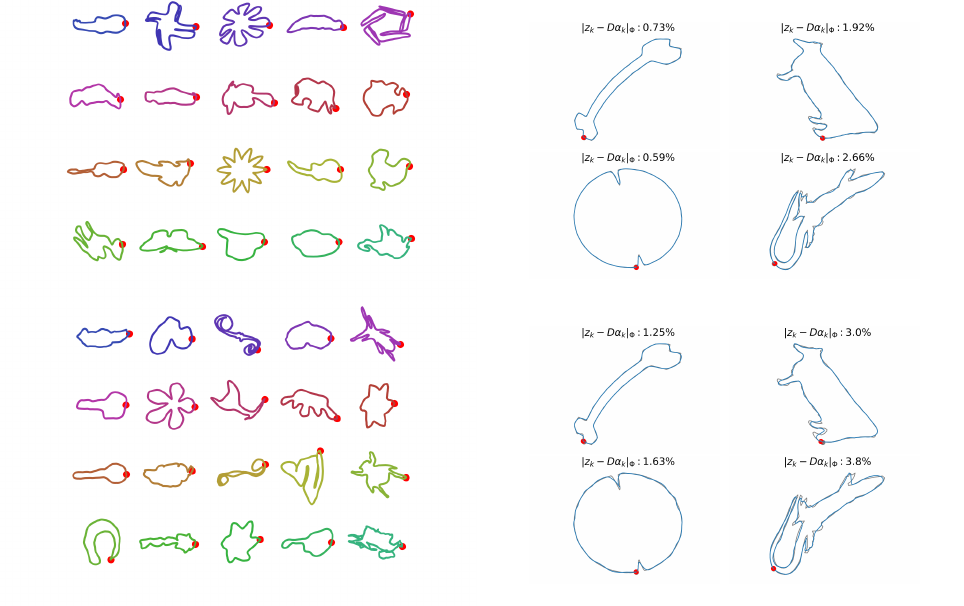}
	\caption{\small\textbf{Dataset 5.} Up: our method. Down: align-first method. Left: first $20$ atoms of the shape dictionary of $J = 230$ atoms, taking $N_0 = 30$ out of them to reconstruct the $K = 14000$ shapes represented as configurations of dimension $N = 200 < J$. Right: four examples of reconstruction (blue) over the original data (gray). The RMSEs are $1.57\%$ and $2.71\%$, respectively.}
	\label{fig:mpeg_synth}
\end{figure}

In Figures \ref{fig:worms}, \ref{fig:6376_open_worms}, \ref{fig:hands}, \ref{fig:leaves_sym}, \ref{fig:leaves_Bsplines} and \ref{fig:mpeg_synth}, we show the dictionary learned by 2DKSD and the align-first method with specific parameters for each dataset (Table \ref{tab:RMSE}), alongside with four examples of reconstructed data superimposed over the original one. In each case, the reconstruction errors
$$|\z_k - \D\alph_k|_\Phh = \frac{|\z_k - \D\alph_k|_\Phh}{|\z_k|_\Phh}$$
are expressed in percentage, and can be understood as an absolute error multiplied by $100$, or as a proportion of $|\z_k| = 1$ as well. Note that, as mentioned in Section \ref{sub:sparse_coding}, the 2DKSD ensures that these errors are equal to $d_F([\z_k],[\D \alph_k])$, which is not the case for the align-first method. We also indicate the root mean square reconstruction errors (RMSE), defined as
$${\displaystyle \sqrt{ \frac{ E}{K}} = \sqrt{ \frac{1}{K} \sum_{k=1}^K \left|\z_k - \D \alph_k \right|_\Phh^2 }} \quad \text{and} \quad {\displaystyle  \sqrt{ \frac{ E'}{K}} = \sqrt{ \frac{1}{K} \sum_{k=1}^K \left|\widetilde{\z}_k - \D' \alph'_k \right|_\Phh^2 } },$$ respectively, and also expressed in percentage. We summarize them in Table \ref{tab:RMSE}, where we specify the parameters used in the examples, as well as a typical run-time\footnote{All experiments were run on a standard laptop with an Intel® Core™ i5-7200U CPU running 4 cores at 2.50GHz, with 7,7 Gb of RAM.}
for $T = 30$ iterations in Algorithm \ref{algo:2DKSD}, which is sufficient for the loss to converge in all datasets, and including the duration of the final sparse coding of the whole dataset. To reduce ORMP run-time for larger datasets (such as Datasets 1 and 2, marked with a star * in the following table), we applied Algorithm \ref{algo:ORMP} in parallel and independently on each data point $\z_k$. To learn overcomplete dictionaries on very big datasets (such as Dataset 5, marked with two stars ** in the table), we chose to update the weights $\alph_k$ stochastically and in parallel by batches of $4096$, after initializing them to zero. Obviously, additional efforts could be done to optimize the implementation and reduce run-time on big datasets, but this falls out of the scope of the present work, and is kept for the future.

\begin{figure}[h]
	\centering
	\includegraphics[clip,trim={.07cm 0 .3cm 0}, width=\linewidth]{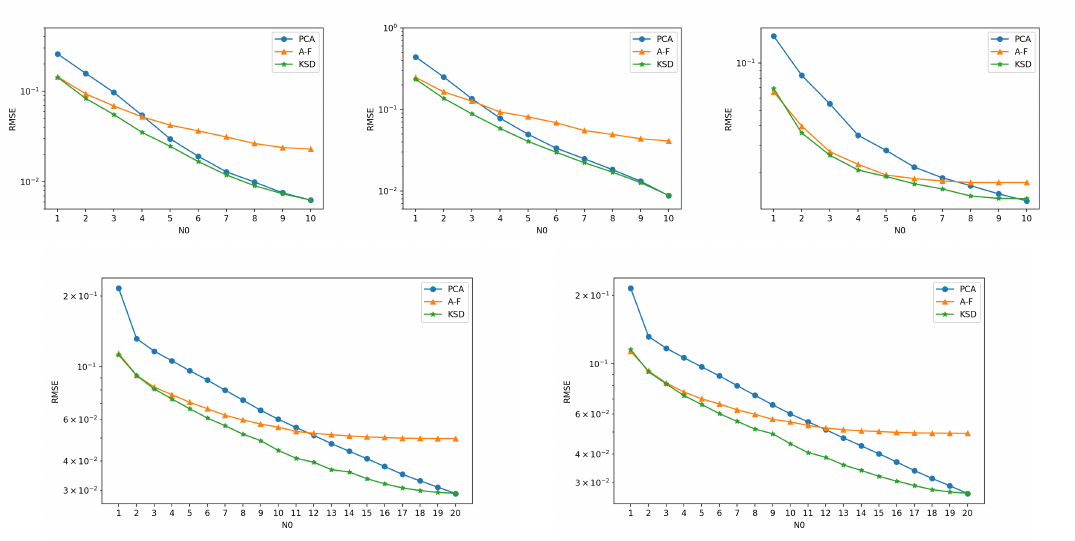}
	\caption{\small\textbf{RMSE of complex PCA, (real) align-first and 2DKSD, in logarithmic scale.} From left to right, top to bottom: Datasets 1 to 4b. In each graph, we plot the RMSE as a function of the number $N_0$ of atoms used in the reconstructions. For complex PCA, we used the first $N_0$ modes. For align-first and 2DKSD, we learned a dictionary parameterized by $(N_0,J)$ for each value of $N_0$, with $J$ set to $10$ (top row) or $20$ (bottom row).}
	\label{fig:error_rates}
\end{figure}

In Figure \ref{fig:error_rates}, we illustrate in logarithmic scale the RMSE obtained with complex PCA \textit{v.s.} (real) align-first \textit{v.s.} 2DKSD, after learning atoms from Datasets 1 to 4b. For varying values of $N_0$, we compare the RMSE of a reconstruction relying on the first $N_0$ PCA modes, and the RMSEs of an align-first or 2DKSD dictionary learned with the parameters $(N_0,J)$. $J \geq N_0$ is fixed to the same values used in the examples shown in the figures, and is such that $J < N$. This comparison is relevant for two reasons. First, the dictionary of PCA atoms is a natural candidate when learning an undercomplete dictionary containing less atoms than the dimension of the configuration space. Second, while complex PCA relies on complex weights but determines the atoms in a specific way, the align-first method is less constrained in the choice of atoms but relies on real weights. Our 2DKSD method is therefore a good compromise between them. We have also explored the $J > N$ case, which leads to an overcomplete dictionary (Dataset 5, Figure \ref{fig:mpeg_synth}).\newline

\begin{table}[h]
	\caption{\textbf{RMSE and parameters used to generate the results of Figures \ref{fig:worms} to \ref{fig:leaves_Bsplines}.}}
	\label{tab:RMSE}
	\centering
	\begin{tabular}{|c|c|c|c|c|c|c|c|}
		\hline
		\textbf{Dataset} & $N$ & $K$ & $N_0$ & $J$ & Indicative run-time & RMSE, \textbf{ours} & RMSE, \textbf{align-first} \\
		\hline
		1 & 20 & >5500 & 5 & 10 & 21 s *& $2.36\%$ & $4.19\%$ \\
		2 & 12 & > 6300 & 5 & 10 & 24 s *& $4.04 \%$ & $7.95 \%$ \\
		3 & 56 & 40 & 3 & 10 & < 0.4 s & $2.48\%$ & $2.73\%$ \\
		4 & 200 & > 1500 & 5 & 20 & 18 s & $6.45\%$ & $7.12\%$ \\
		4b & 40 & > 1500 & 5 & 20 & 12 s & $6.41\%$ & $7.01\%$\\
		5 & 200 & 14000 & 30 & 230 & 441 s ** & $1.57\%$ & $2.71\%$ \\
		\hline
	\end{tabular}
\end{table}

\paragraph{Comparison with align-first and complex PCA}
We argue that the align-first method provides less satisfying results than our shape dictionary, both mathematically and numerically. Mathematically, the optimized loss functional \eqref{eq:align_first}, denoted by $E'$, is necessarily larger than our loss $E$ \eqref{eq:loss} since weights are constrained to lie on the real line. Intuitively, they can scale but not rotate the atoms before summing them. In contrast, the complex weights in our method scale and rotate the atoms, resulting in a smaller loss $E$. Numerically, this leads to a visible difference in the RMSEs, as seen in Figure \ref{fig:error_rates}. In particular, the loss obtained with our algorithm $E$ can be a third or a fourth of the one resulting from align-first $E'$, as in Figures \ref{fig:worms} and \ref{fig:6376_open_worms}.
Besides, the dictionary computed by align-first often contains distorted, irregular or unrealistic atoms, which reflect the effort required to fit the data, in spite of the pre-alignment. When analyzing datasets of shapes, it is sometimes preferable to obtain realistic atoms and visually accurate reconstructions instead of perfect reconstructions but unrealistic atoms. Visually realistic atoms indeed offer a way to hypothesize on the nature of the variability of the dataset.

\begin{figure}[H]
	\centering
	\includegraphics[width=\linewidth]{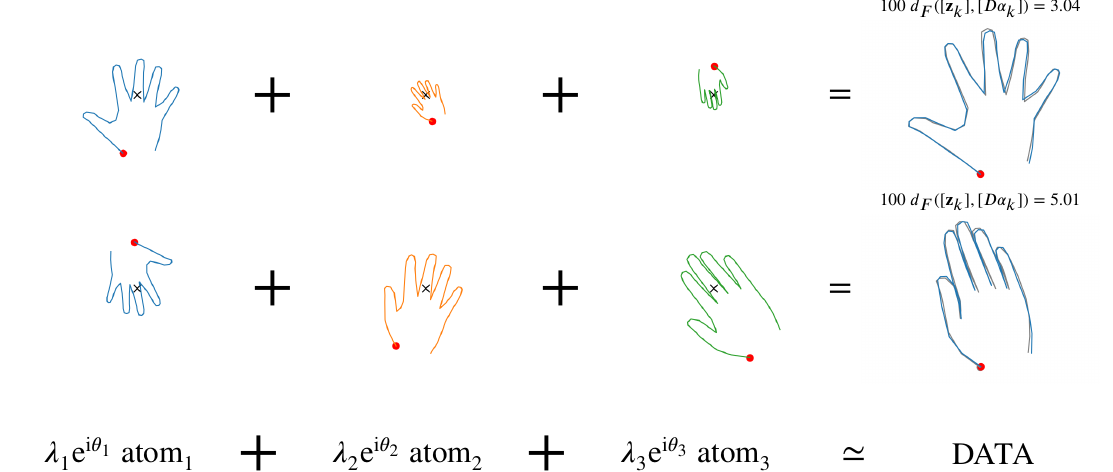}
	\caption{\small \textbf{Complex linear combinations offer more freedom in the reconstruction.} Left: three atoms weighted with different complex coefficients. Right: their sum (blue) superimposed over reconstructed data (gray). Black crosses indicate the origin.}
	\label{fig:complex_combination}
\end{figure}

Figure \ref{fig:error_rates} demonstrates the benefits of using complex numbers in the linear combinations. For increasing values of $N_0$, the align-first RMSE indeed becomes significantly larger even when compared to that of complex PCA. As another illustration of their relevance, Figure \ref{fig:complex_combination} features two different weighted sums of the same three atoms that allow reconstructing very dissimilar shapes. It is not surprising for the first hand to be correctly reconstructed as, at first order, its shape is similar to that of the first atom, which is weighted by the coefficient of largest magnitude. The remainder is compensated by the other two atoms. For the second hand, however, it is more surprising to see that hands with open fingers, scaled, rotated, and then summed together, manage to produce a hand with closed fingers.

Last, Figure \ref{fig:error_rates} shows that 2DKSD is more appropriate than complex PCA for $N_0 < J$. When learning a KSD dictionary that uses all the $N_0 = J < N$ atoms to reconstruct data, the RMSE is close to that of complex PCA. In fact, the minimal RMSEs should mathematically be equal whenever the average configuration of the dataset corresponds to the zero configuration, or if it is not discarded from the dataset (see the details of complex PCA discussed above). Indeed, both solutions find the $J$-dimensional subspace that is the closest to the data points, \textit{i.e.}, the subspace spanned by the $J$ PCA modes.\newline

\paragraph{Comments on the examples comparing 2DKSD and align-first}
As is seen in Figures \ref{fig:worms} and \ref{fig:6376_open_worms}, nematode shapes seem to be efficiently reconstructed using not more than $5$ atoms. For both datasets 1 and 2, the reconstructions are visually satisfying and numerically accurate. The results for Dataset 2 are remarkable in that, although the original shapes do not have the characteristic smoothness of freely moving worms, the algorithm has less difficulty in the reconstruction than align-first. Moreover, it produces realistic atoms, whose curve parameterization is well-balanced, in contrast to those of align-first where consecutive control points are not regularly distributed along the curve.

In Figure \ref{fig:hands} reporting the results for Dataset 3, both methods result in very similar losses, but the atoms are less realistic with align-first. They exhibit self-intersecting and stretched fingers. In this case, it is an informative feature as it indicates to which extent fingers ``tend'' to be closed. Our algorithm is then an alternative to the align-first method.

The results in Figures \ref{fig:leaves_sym} and \ref{fig:leaves_Bsplines} on the leaf datasets 4 and 4b are similar regardless of the representation (landmarks or B-splines) used. Our loss $E$ is less than $84\%$ of the comparison loss $E'$, and we obtain a dictionary with realistic leaves, while align-first contains twisted shapes.

Finally, we observe that the overcomplete dictionaries learned by both methods on Dataset 5, using $20$ of their atoms shown in Figure \ref{fig:mpeg_synth}, look fairly similar. Thus, 2DKSD offers no particular advantage over align-first regarding the visual realisticness of the atoms in that case, but leads to a significantly smaller RMSE. It also reproduces better high-frequency features of the silhouettes, as in the cow and lizard shapes.

\section{Conclusion}
\label{sec:ccl}

 Our 2D Kendall Shape Dictionary approach is a natural adaptation of usual learning techniques to Kendall's nonlinear manifold and does not require sophisticated operations. For the classical full Procrustes metric, it simplifies to a nearly standard formulation in which complex weights are used when combining the atoms of the dictionary. As our main contribution, this formulation has the double property to enjoy \textit{both a strong theoretical justification and a simple algorithmic framework}.
 We have demonstrated the positive impact of using a complex setting: datasets do not require to be aligned along a reference mean because atoms are freely scaled and rotated before being summed to reconstruct shapes independently. This flexibility increases reconstruction accuracy and allows dictionary atoms to remain visually realistic. We have also extended Kendall's space of planar shapes, initially defined for discrete configurations of landmarks, to continuously-defined interpolating curves, by introducing the general notion of configuration. Hopefully, our method is a promising tool for characterizing complex phenotypes from biological images.

\section*{Acknowledgments}
A.S. is especially thankful to Jean Feydy for drawing her attention to reference \cite{mairal2014}. She is grateful for the support from \'Ecole Normale Sup\'erieure and \'Ecole polytechnique f\'ed\'erale de Lausanne. V.U., J.F., and M.U. were supported by the Swiss National Science Foundation under Grant \verb|200020_162343 / 1|. V.U. was also partly supported by EMBL core fundings. J.F. was also partly supported by the Swiss National Science Foundation with grant agreement \verb|P2ELP2_181759|.

\section{Appendix}
\label{sec:appendix}

\subsection{Hermitian products hold twice more information than scalar products}
\label{sub:appendix_hermitian}
In the sequel, we make the identification $\C^N \simeq \R^{2N}$ through the mapping $\z \mapsto z = (\Re(\z),\Im(\z))^T$, where $\Re(\z) = (\Re(\z[0]),...,\Re(\z[N-1]))$ and likewise for the imaginary part. We use bold and italic to distinguish the complex and real counterparts of the vectors. If $(\C^N,\Phh)$ is endowed with a Hermitian inner product $(\z,\w) \mapsto (\z ~|~ \w)_\Phh := \z^*\Phh \w$ associated to the matrix $\Phh \in \C^{N \times N}$, then one can define on $\R^{2N}$ the scalar product $\langle \cdot, \cdot \rangle_\Phh$ canonically associated to it as
\begin{equation}
\langle z, w \rangle_\Phh := \Re \left( (\z ~|~ \w)_\Phh \right),
\end{equation}
defining the same norm $\langle z, z \rangle_\Phh = (\z ~|~ \z)_\Phh.$
Due to the sesquilinearity of the Hermitian product, we have that $ (\mathrm{i} \z ~|~ \w)_\Phh = -\mathrm{i} (\z ~|~ \w)_\Phh$, so that $ \Re \left( (\mathrm{i} \z ~|~ \w)_\Phh \right) = \Im \left( (\z ~|~ \w)_\Phh \right) $, and then
\begin{equation}(\z ~|~ \w)_\Phh =  \langle z, w \rangle_\Phh + \mathrm{i} \langle R_{\pi/2} ~\odot~ z, w \rangle_\Phh,\end{equation}
where by $R_\theta \odot z = \cos \theta ~ (\Re(\z),\Im(\z))^T + \sin \theta ~ (-\Im(\z),\Re(\z))^T$ we denote the image (in the real setting) of $\z$ by a rotation by $\theta$, namely, the real counterpart of $e^{\mathrm{i} \theta} \z$. Here, $R_{\pi/2} ~\odot~ z = (-\Im(\z),\Re(\z))^T$.

As a consequence, $(\z ~|~ \w)_\Phh = 0$ if and only if $\langle z, w \rangle_\Phh = 0$ and $\langle R_{\pi/2} ~\odot~ z, w \rangle_\Phh = 0$. In fact, a stronger property can be deduced: $\forall \theta \in [0,2\pi)$, $\langle R_{\theta} ~\odot~ z, w \rangle_\Phh = \langle z, R_{\theta} ~\odot~ w \rangle_\Phh = 0$.
This is because $R_{\theta} ~\odot~ z = \cos \theta ~z + \sin\theta ~R_{\pi/2} ~\odot~ z$, and we conclude by bilinearity of the scalar product, and then by exchanging the roles of $z$ and $w$.
In other words, \textit{Hermitian orthogonality can be understood as a real orthogonality between any rotated image of $\z$ and $\w$ (and conversely)}.

\begin{remark}
Suppose that $\Phh = \mathbf{A} + \mathrm{i} \mathbf{B}$ with $\mathbf{A},\mathbf{B} \in \R^{N \times N}$.
The scalar product is computed as $\langle z, w \rangle_\Phh = z^T \mathbf{\Psi} w,$
with $\mathbf{\Psi} = \begin{pmatrix}
	\mathbf{A} & -\mathbf{B}\\
	\mathbf{B} & \mathbf{A}
\end{pmatrix}$.
\end{remark}

\subsection{Two lemmas}
\label{appendix:two_lemmas}
In this section, we establish two lemmas. The first one is used in the proof of Proposition \ref{prop:simple}, and the second one states that the dictionary update \eqref{eq:dico_update} does not involve the matrix $\Phh$.

\begin{lemma}
	\label{lm:2_pbs}
	Let $\hat{\alph} \in \C^J$ be a minimizer of $\min \limits_{|\alph|_0 \leq N_0} |\z - \D \alph|_\Phh.$
	If $\D\hat{\alph} \neq \mathbf{0}$, then any term in \eqref{eq:dico_elementary} is also minimized by $\hat{\alph}$.
	Otherwise, $\z$ is orthogonal to the columns of $\D$ and the distances in \eqref{eq:dico_elementary} are all maximized to $1$. In that case, any $\widetilde{\alph}$ satisfying $\D \widetilde{\alph} \neq \mathbf{0}$ is a minimizer of any term in \eqref{eq:dico_elementary}.
\end{lemma}
\begin{proof}
	The quantity $|\z - \D \alph|_\Phh$ reaches a minimum in $\{\alph ~|~ |\alph|_0 \leq N_0 \}$. It corresponds to the minimal distance to $\z$ of a subspace generated by at most $N_0$ atoms $\d_j$ (see previous discussions about sparse coding). Let $\hat{\alph}$ denote the corresponding minimizer. It satisfies $\min \limits_{|\alph|_0 \leq N_0} |\z - \D \alph|_\Phh = |\z - \D \hat{\alph}|_\Phh$.
	By definition of a projection, $ |\z - \D \hat{\alph}| \geq |\z - \mathrm{P}_{\C \D \hat{\alph}}\z|$. Moreover, if $\lambda \in \C$ is such that $\mathrm{P}_{\C \D \hat{\alph}} \z = \lambda \D \hat{\alph}$, then $|\z - \mathrm{P}_{\C \D \hat\alph}\z| = |\z − \D(\lambda \hat\alph)|$, where $|\lambda\hat\alph|_0 \leq N_0$. The last term is then greater or equal to $\min \limits_{|\alph|_0 \leq N_0} |\z - \D \alph|_\Phh$, hence it comes that
	\begin{equation} \label{eq:above}
	|\z - \D \hat{\alph}|_\Phh = |\z - \mathrm{P}_{\C \D \hat{\alph}}\z|_\Phh.
	\end{equation}
	Using the same arguments, one easily shows that this also corresponds to $\min \limits_{|\alph|_0 \leq N_0} |\z - \mathrm{P}_{\C \D \alph}\z|_\Phh$.
	As a useful remark, notice that the equality \eqref{eq:above} implies that $\mathrm{P}_{\C \D \hat{\alph}}\z = \D \hat{\alph}$.
	
	Coming back to the original equalities \eqref{eq:dico_elementary}, if $\D \hat{\alph} \neq \mathbf{0}$, then all terms are minimized by $\hat{\alph}$. Otherwise, if $\D \hat{\alph} = \mathbf{0}$, then $\z$ is orthogonal to the vectors $\d_j$, and $\z$ is not correlated to any shape generated by the $\d_j$, inducing a maximal distance $d_F$. In particular, any $\widetilde{\alph}$ such that $|\widetilde{\alph}|_0 \leq N_0$ and $\D \widetilde{\alph} \neq \mathbf{0}$, can be taken as a minimizer of any term in \eqref{eq:dico_elementary}.
\end{proof}

\begin{lemma}\label{lm:dico_update}
$E$ being defined in \eqref{eq:loss}, a solution to $\min \limits_{\D \in \C^{n \times J}} E(\D,\A)$, for a fixed value of $\A$, is given by \eqref{eq:dico_update}, and its expression is independent from the matrix $\Phh$.
\end{lemma}
\begin{proof}
We re-write the problem into a standard $\ell^2$ form:
\begin{align*}
\min\limits_{\D} E(\D,\A) = \min\limits_{\D} \sum\limits_{k = 1}^K |\D \alpha_k - \z_k|_\Phh^2 & = \min\limits_{\D} \sum\limits_k |\sqrt{\Phh} \D \alpha_k - \sqrt{\Phh} \z_k|^2 \\
& = \min\limits_{\D} \| \sqrt{\Phh}(\D \A - \Zbf) \|_{\text{Fro}}^2 \\
& = \min\limits_{\D} \| (\A^* \D^* - \Zbf^*)\sqrt{\Phh} \|_{\text{Fro}}^2 \\
& = \min\limits_{\mathbf{H}} \sum\limits_{i=1}^n |\A^* \mathbf{h}_i - \mathbf{g}_i|^2,
\end{align*}
$\mathbf{h}_i$ and $\mathbf{g}_i$ denoting the columns of $\mathbf{H} = \D^* \sqrt{\Phh}$ and $\mathbf{G} = \Zbf^* \sqrt{\Phh}$. It is known that the elementary problems are solved by $\mathbf{\hat{h}}_i = (\A^*)^+ \mathbf{g}_i$. This gives $\mathbf{\hat{H}}^* = \mathbf{G}^* \A^+$ and finally $\hat{\D} = \Zbf \A^+$, after inverting with $(\sqrt{\Phh})^{-1}$.
\end{proof}

\subsection{Mean shape}
\label{sub:appendix_mean}

\begin{figure}[h]
	\centering
	\includegraphics[width=\linewidth]{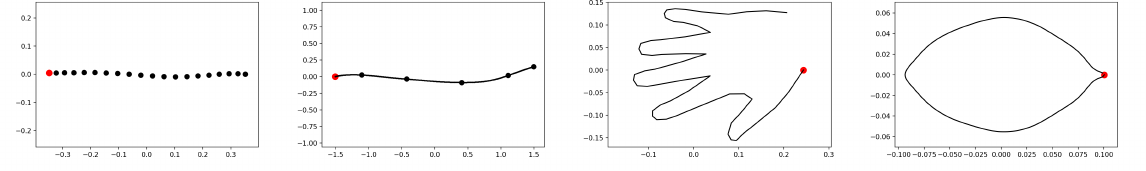}
	\caption{\small\textbf{Fréchet mean shape of Datasets 1, 2, 3 and 4 with respect to} $\mathrm{dist} = d_F$. We display one representative pre-shape, up to rotations.}
\label{fig:8}
\end{figure}

The mean shape of a dataset can be defined as the Fréchet mean of the points $[\z_k]$ scattered on Kendall's manifold \cite{dryden2016}, with respect to one of the three distances $d_F$, $d_P$, or $\rho$. It is the unique global minimizer $\z_\mathrm{mean}$ of
$$\sum_{k=1}^K \mathrm{dist}([\z_\mathrm{mean}],[\z_k])^2,$$
when it exists. The Fréchet mean with respect to the full distance $\mathrm{dist} = d_F$ (see Figure \ref{fig:8}) can be found as the shape of the eigenvector associated to the greatest eigenvalue of the operator \cite{dryden2016,srikla2016}
$$\sum_k \mathrm{P}_{\C \z_k},$$
where $\z_k \in \S$ are pre-shapes, and where $\mathrm{P}_{\C \z_k} = \z_k \z_k^* \Phh$ is the orthogonal projector onto the complex vector line generated by $\z_k$ relatively to the Hermitian product $\Phh$. To prove it, let us consider the problem in the pre-shape sphere. We obtain:
\begin{equation}
\argmin \limits_{\z \in \S} \sum |\z - \mathrm{P}_{\C \z_k} \z|_\Phh^2 = \argmax \limits_{\z \in \S} \sum |\mathrm{P}_{\C \z_k} \z|^2_\Phh = \argmax \limits_{\z \in \S} \z^* \Phh \sum_k \mathrm{P}_{\C \z_k} \z,
\end{equation}
where we used that $|\z - \mathrm{P}_{\C \z_k} \z|_\Phh^2 = 1 - |\mathrm{P}_{\C \z_k} \z|^2_\Phh$.
A similar proof for landmarks ($\Phh = \mathrm{Id}$) can be found in \cite[p. 178]{dryden2016}.\newline

\paragraph{Link with the mean-shape curve of \cite{schmitter2018}}
In this article related to ours, a \textit{mean-shape curve} of a family of curves $\{r_1,...,r_K\}$ in $H := \L^2([0,1],\R^2)$ endowed with the usual norm is defined as an optimal curve
\begin{equation}
r_\mathrm{mean} \in \argmin \limits_{ \substack{|r|_H = 1 \\ \bar{r} = 0 } } \sum_{k=1}^K |r - \mathrm{P}_k r|^2_H = \argmax \limits_{ \substack{|r|_H = 1 \\ \bar{r} = 0 } } \sum_{k=1}^K |\mathrm{P}_k r|^2_{H},
\end{equation}
where $\mathrm{P}_k$ is the \textit{similarity projector}, \textit{i.e.}, the orthogonal projection onto the subspace $S_{r_k}$ of dimension $4$ associated to $r_k$ containing all the images up to similitude transforms of $r_k$. It writes

\begin{align}
S_{r} & = \left\{\lambda \begin{pmatrix}
\cos \theta & - \sin \theta \\
\sin \theta & \cos \theta
\end{pmatrix} r + \begin{pmatrix}
\alpha\\
\beta
\end{pmatrix} ~|~ \lambda \in \R, \theta \in [0,2\pi), \alpha, \beta \in \R \right\} \\
& =
\R \left\{ \begin{pmatrix} r^x\\r^y \end{pmatrix}, 
\begin{pmatrix} -r^y\\r^x \end{pmatrix},
\begin{pmatrix} 1 \\ 0 \end{pmatrix},
\begin{pmatrix} 0\\ 1 \end{pmatrix}  \right\}.
\end{align}
It can be shown that $r_\mathrm{mean}$ belongs to the eigenspace associated to the \textit{second} greatest eigenvalue of $\sum_k \mathrm{P}_k$ as
\begin{equation}
\sum_{k=1}^K \mathrm{P}_k r_\mathrm{mean} = \lambda_2 r_\mathrm{mean}, \qquad |r_\mathrm{mean}| = 1
\end{equation}
and, as a consequence, the mean-shape curve of interpolating curves is also an interpolating curve (as soon as constants are themselves interpolating curves), since $r_\mathrm{mean} \in \text{Im}(\sum \mathrm{P}_k)$. Aligning the dataset to the mean-shape curve then consists in taking
$$\widetilde{r_k} := \mathrm{P}_k r_\mathrm{mean}.$$

In fact, in the complex setting, if $r_1,...,r_K \in \L^2([0,1],\C)$ are centered and normalized, the orbit of the mean-shape interpolating curve $[r_\mathrm{mean}]$ up to rotations \textit{is then exactly the Fréchet mean} of the orbits $[r_1],...,[r_K]$ in the curve counterpart of the shape space with respect to the full distance $d_F$. In other words, if $\z_1,...,\z_K$ are the control vectors of $r_1,...,r_K$, then the control vector $\z_\mathrm{mean} \in \S$ of $r_\mathrm{mean}$ is one representative, up to rotations, of the Fréchet mean of $[\z_1],...,[\z_K]$. Also, the notion of alignment proposed in \cite{schmitter2018} coincides with that involved in the full distance (see \eqref{eq:d_F_proj}), if working with pre-shapes.\newline

\subsection{B-spline curves}

\label{sub:appendix_splines}

We find it is worthwhile to present our setting for B-splines, which are popular tools to construct interpolating curves (\cite{unser1993,brigger2000}). We provide explicit expressions for the basis functions $\phi_n$. Spline curves (interpolating curves, more generally) motivate the introduction of a Hermitian product $\Phh$ and a notion of shift configuration $\u$, for which projecting on the subspace orthogonal to $\u$ is akin to centering the spline curve. Readers can rely on this example to adapt our framework to other representations, such as Hermite-spline curves. 

B-spline curves have parameters called \textit{control points} that are regularly spaced along a continuous parameter $t$. A B-spline with $M$ control points holds $N = M$ degrees of freedom, as in
\begin{equation}
\forall t \in [0,1], \quad r(t) = \sum_{n = 0}^{M-1} \z[n] \phi_n(t),
\end{equation}
where the basis functions $\phi_n$ can be obtained, for instance, from the cubic B-spline generator function
\begin{equation}
\beta^3(t) = \left\{
\begin{array}{cc}
|t|^3/2 - t^2 + 2/3 & \text{for } 0 \leq |t| \leq 1 \\
\hskip-.9cm (2 - |t|)^3/6 & \text{for } 1 \leq |t| \leq 2 \\
\hskip -2.5cm 0 & \text{otherwise.} \\
\end{array} \right.
\end{equation}

Closed cubic B-splines curves are then obtained with
\begin{align}
\text{for } n = 0, \forall t \in [0,1], \qquad &\phi_0(t) = \beta^3(Mt) + \beta^3(Mt-M),\\
\text{for } n = 1, \forall t \in [0,1], \qquad &\phi_1(t) = \beta^3(Mt - 1) + \beta^3(Mt-1-M),\\
\forall n = 2,...,M-2, \forall t \in [0,1], \qquad &\phi_n(t) = \beta^3(Mt-n), \\
\text{for } n = M-1, \forall t \in [0,1], \qquad &\phi_{M-1}(t) = \beta^3(Mt + 1) + \beta^3(Mt + 1 - M).
\end{align}

One can check that $\mathbbm{1}$ is a cubic B-spline and that it corresponds to the shift configuration $\u = (1,...,1)$.
We know that the temporal mean of the spline curve, defined after Definition \ref{df:centre}, coincides with the product $\frac{\u^* \Phh \z }{|\u|_\Phh^2} = \u^* \Phh \z = \overline{\z^* \Phh \u}$.
In the case of closed B-splines, it is simply the usual arithmetic mean of the control points
\begin{equation}
\bar{r} = \frac{1}{M} \sum_{m = 0}^{M-1} \z[m],
\end{equation}
which is subtracted to $\z$ when centering the configuration. Please note that we have more sophisticated expressions for closed and open Hermite-spline curves, that we detail in a future work.

\subsection{Riemannian structure of the shape space}
\label{sub:appendix_riem}
We provide a concise description of the Riemannian structure of the shape space $\Sig$ (our method does not use it in a crucial way). Using the same conventions as in Appendix \ref{sub:appendix_hermitian}, we use bold and italic to denote the complex and real version of a vector by $\z \in \C^N$ and $z \in \R^{2N}$, respectively. The real inner product $(z,w) \mapsto \Re(\z^* \Phh \w)$ canonically associated to the Hermitian inner product $\Phh$ is denoted as $\langle\cdot, \cdot \rangle_\Phh$. The shape space based on $(\R^{2N},\langle\cdot, \cdot \rangle_\Phh)$ has a Riemannian structure that is similar to the usual complex projective space $\C\mathbb{P}^{N-2}$. The latter is the shape space based on $(\R^{2N},\langle\cdot, \cdot \rangle_{\mathrm{can}})$, defined for landmarks and the standard inner product \cite{kendall1977,dryden2016,srikla2016,gallot2004,gallier2011}.

The Riemannian structure of $\Sig$ is inherited from that of $\S$, a sphere in finite dimension whose structure is well understood.
Being a smooth submanifold of $(\R^{2N},\langle\cdot, \cdot \rangle_\Phh)$, $\S$ is endowed with the induced Riemannian metric. The tangent space to $\S$ at $z$ is
\begin{equation} \label{eq:tangent}
T_z(\S) = \{v \in \R^{2N} ~|~ \langle z, v \rangle_\Phh = 0 \},
\end{equation}
which is equipped with the Riemannian inner product $\forall v_1,v_2 \in T_z \S, \langle v_1, v_2 \rangle_z := \langle v_1, v_2 \rangle_\Phh $.

The space space $\Sig$ is the quotient of $\S$ by the group of planar rotations $U(1)$. The action of $U(1)$ on the Riemannian manifold $\S$ is smooth, free, and proper on $\S$, with $U(1)$ being a Lie group acting by isometries, meaning that
\begin{equation}
\forall \theta \in U(1), \langle v_1, v_2 \rangle_z = \langle \mathrm{R}_{\theta} ~\odot~ v_1, \mathrm{R}_{\theta} ~\odot~ v_2 \rangle_{ \mathrm{R}_{\theta} ~\odot~  z}.
\end{equation}
The resulting quotient $\Sig$ consequently inherits a (unique) Riemannian structure on $\Sig = \S / U(1)$ such that
$\mathbf{\Pi} : \left( \begin{array}{ccc}
\S & \to & \Sig \\
z & \mapsto & [\z]
\end{array} \right)$
is a Riemannian submersion \cite{gallot2004,gallier2011}. 
This structure can be described as follows \cite{gallot2004,gallier2011,srikla2016}. 
At each point $z$ of $\S$, the tangent space $T_z(\S)$ can be split into two subspaces orthogonal to each other with respect to $\langle\cdot, \cdot \rangle_\Phh$ as
\begin{equation}
T_z(\S) = T_z([z]) \overset{\perp}{\oplus} H_z(\S),
\end{equation}
where the \textit{vertical subspace} $T_z([z]) = \mathrm{Ker}(d\pi_z) = \R (R_{\pi/2} ~\odot~ z)$ is the space tangent at $z$ to the orbit $[z]$, while the orthogonal complement is the \textit{horizontal subspace}
\begin{equation}
H_z(\S) = \{v \in \R^{2N} ~|~ \langle z, v \rangle_\Phh = 0 \text{ and } \langle z, R_{\pi/2} ~\odot~ v \rangle_\Phh = 0 \} \simeq \{\v \in \C^N ~|~ \z^* \Phh \v = 0\}.
\end{equation}
The last identification uses the properties of the Hermitian inner product seen in Appendix \ref{sub:appendix_hermitian}. By definition, the tangent space $T_{[z]}(\Sig)$ is then identified to the horizontal subspace $H_z(\S)$ through the isometry $d\pi_z : v \in H_z(\S) \mapsto [v] \in T_{[z]}(\Sig)$.
The Riemannian inner product on $T_{[z]}(\Sig)$ is then obtained as
\begin{equation}
\langle \langle [v_1], [v_2] \rangle \rangle_{[z]} := \langle v_1, v_2 \rangle_z,
\end{equation}
which does not depend on the particular choice of the representative pre-shape $z$.

Geodesics and Riemannian distance $\rho$ (Definition \ref{df:rho}) can then be defined on $\Sig$. The Riemannian distance $\rho_\S$ on $\S$ is equal to the length of the shortest path joining $z$ to $w$,
\begin{equation}
\rho_{\S}(z,w) = \arccos \langle z, w \rangle_\Phh.
\end{equation}
The corresponding shortest geodesic arc $\{\alph(t)\}$ that joins $z$ to $w$ is
\begin{equation} \label{eq:geodesic}
\forall t \in [0,1], \quad \alph(t) = \frac{1}{\sin r}(\sin((1-t)r) z + \sin(t~r)w), \quad \text{where } r = \rho_{\S}(z,w).
\end{equation}
The shortest geodesic path between the shapes $[\z]$ and $[\w]$ is the quotient path $\{[\tilde{\alph}(t)]\}$, where $\tilde{\alph}$ is the geodesic in $\S$ that joins $\tilde{z} \in [\z]$ to $w$, where $\tilde{z}$ is chosen so that the corresponding (complex) pre-shape $\tilde{\z}$ is optimally rotated along $\w$.
The Riemannian distance in the shape space is then the length of $\tilde{\alph}(t)$, computed as $\arccos \langle \tilde{z}, w \rangle_\Phh = \arccos (\Re (\tilde{\z}^* \Phh \w))$. It also corresponds to the geodesic distance between $\tilde{z}$ and the set $[\w]$. From an argument in the proof of Proposition \ref{prop:link_d_P_d_F}, we know that, since $\tilde{\z}$ is optimally rotated, it is also equal to $\arccos |\tilde{\z}^* \Phh \w | = \arccos |\z^* \Phh \w|$, thus resulting in the equality of Definition \ref{df:rho}.

\bibliographystyle{siamme}
\bibliography{article_refs}

\end{document}